\documentclass[superscriptaddress,aps,longbibliography,amsmath,amssymb,nofootinbib]{revtex4-2}
\usepackage{empheq} 
\usepackage{float}
\usepackage{placeins}
\usepackage{multirow}
\usepackage{xcolor}
\usepackage{graphicx}
\usepackage{subcaption}
\usepackage[toc,page]{appendix}

\def\mat#1{\pmb{#1}}
\def\vector#1{\boldsymbol{#1}}
\def\sym#1{\mbox{\textit{#1}}}
\def\Re{\mbox{\textit{Re}}}

\def\SNR{\mbox{\textit{SNR}}}
\def\loss{\mathcal{L}}
\def\lossClassic{\loss^c}
\def\lossStrict{\loss^s}
\def\lossMean{\loss^m}
\DeclareMathOperator*{\argmin}{arg\,min}

\def\LW#1{\dimexpr#1\linewidth-.5em}

\begin{document}
\title{Reconstructing unsteady flows from sparse, noisy measurements with a physics-constrained convolutional neural network}
\author{Yaxin Mo}
\affiliation{Department of Aeronautics, Imperial College London, SW7 2AZ, UK}
\author{Luca Magri}
\email{l.magri@imperial.ac.uk}
\affiliation{Department of Aeronautics, Imperial College London, SW7 2AZ, UK}
\affiliation{The Alan Turing Institute, London, NW1 2DB, UK}

\maketitle

\section*{Abstract}
Data from fluid flow measurements are typically sparse, noisy, and heterogeneous, often from mixed pressure and velocity measurements, resulting in incomplete datasets. 
In this paper, we develop a physics-constrained convolutional neural network, which is a deterministic tool, to reconstruct the full flow field from incomplete data. 
We explore three loss functions, both from machine learning literature and newly proposed: (i) the softly-constrained loss, which allows the prediction to take any value; (ii) the snapshot-enforced loss, which constrains the prediction at the sensor locations; and (iii) the mean-enforced loss, which constrains the mean of the prediction at the sensor locations. 
The proposed methods do not require the full flow field during training, making it suitable for reconstruction from incomplete data. 
We apply the method to reconstruct a laminar wake of a bluff body and a turbulent Kolmogorov flow. 
First, we assume that measurements are not noisy and reconstruct both the laminar wake and the Kolmogorov flow from sensors located at fewer than 1\% of all grid points. 
The snapshot-enforced loss reduces the reconstruction error of the Kolmogorov flow by $\sim 25\%$ compared to the softly-constrained loss.
Second, we assume that measurements are noisy and propose the mean-enforced loss to reconstruct the laminar wake and the Kolmogorov flow at three different signal-to-noise ratios.
We find that, across the ratios tested, the loss functions with harder constraints are more robust to both the random initialization of the networks and the noise levels in the measurements.
At high noise levels, the mean-enforced loss can recover the instantaneous snapshots accurately, making it the suitable choice when reconstructing flows from data corrupted with an unknown amount of noise.
The proposed method opens opportunities for physical flow reconstruction from sparse, noisy data.

\newpage

\section{Introduction}

In many flow applications, particularly experiments, it is useful to estimate a flow field from incomplete data.
Neural networks, with their ability to handle a large amount of data and approximate nonlinear functions, are a principled choice for performing flow reconstruction, which often involves estimating the velocity and/or pressure from sparse measurements of the flow field.

Many network architectures which require ground-truth flow fields during training have been developed for flow reconstruction.
\cite{erichson2020ShallowNeuralNetworks} reconstructs a bluff body wake, ocean temperature and isotropic turbulence from partial observations with a multi-layer perceptron (MLP) network.
\cite{maulikromit2020ProbabilisticNeuralNetworks} reconstructs the ocean temperature, a bluff body wake and an aerofoil with a mixture density network.
Generative adversarial networks have been used to reconstruct rotating turbulence \cite{li2023MultiscaleReconstructionTurbulent,buzzicotti2021ReconstructionTurbulentData} and channel flows \cite{kim2021UnsupervisedDeepLearning,yousif2023DeeplearningApproachReconstructing}.
Other generative networks have also been used to recover turbulence statistics~\cite{sardar2023SpectrallyDecomposedDiffusion}.
Convolutions are a common feature amongst the networks used for reconstruction \cite{matsuo2021SupervisedConvolutionalNetwork,fukami2019SuperresolutionReconstructionTurbulent,fukami2021MachinelearningbasedSpatiotemporalSuper,ozbay2022DeepLearningFluid, guemes2022SuperresolutionGenerativeAdversarial}.

Fewer works are available when ground truth data (i.e., the full flow field) is not available during the training of the networks. \citep{guemes2022SuperresolutionGenerativeAdversarial} reconstructed turbulent flows from noisy point measurements of the flow, in which the points are randomly placed and moving using a generative adversarial network. Their network requires the point measurements to change locations snapshot-to-snapshot, making it suitable for reconstructing flows from results of particle image velocimetry where a large amount of moving point measurements are generated.

When the ground truth is unavailable and measurements are sparse, incorporating prior knowledge into machine learning allows the reconstruction of flow fields from limited measurements.
Physics-informed neural networks (PINNs) can infer the unknowns at unseen locations and have been employed to reconstruct a range of different flows \citep{raissi2019PhysicsinformedNeuralNetworks,raissi2020HiddenFluidMechanics,arzani2021UncoveringNearwallBlood,zhang2021ThreedimensionalSpatiotemporalWind,eivazi2024PhysicsinformedDeeplearningApplications}.
Convolutional neural networks (CNNs) allow an image-based approach, which is suitable for reconstructing high-dimensional flows. 
By incorporating physical knowledge such as the Navier-Stokes equations, CNNs have been employed to reconstruct steady flows \citep{gao2021SuperresolutionDenoisingFluid} and turbulent flows \citep{kelshaw2022PhysicsInformedCNNsSuperResolution, subramaniam2020TurbulenceEnrichmentUsing} from limited measurements.

Depending on the level of understanding that we have of the problem at hand, there are different ways to incorporate physics into training.
Hard-coded constraints such as using $sine$ activation function or Fourier modes are used to ensure periodicity in the output \citep{wong2022LearningSinusoidalSpaces,raynaud2022ModalPINNExtensionPhysicsinformed,ozan2023HardConstrainedNeuralNetowrks}.
However, the most common way to enforce physics is to introduce the governing equations as a penalty term in the loss function to be minimized \citep{raissi2019PhysicsinformedNeuralNetworks,raissi2020HiddenFluidMechanics,gao2021SuperresolutionDenoisingFluid,fukami2023SuperresolutionAnalysisMachine,kelshaw2022PhysicsInformedCNNsSuperResolution}.
In most physics-constrained neural networks, the loss function to be minimized in training is a combination of data loss and physics loss, taking the form
\begin{equation*}
    \text{Loss}_{network} = \lambda_{data}\text{Loss}_{data} + \lambda_{equation}\text{Loss}_{equation},
\end{equation*}
where $\lambda$ is the coefficient, the data loss, $\text{Loss}_{data}$, is a function of only prediction and observations and physics loss, $\text{Loss}_{equation}$, is a function of the prediction derived from the governing equation \citep{fukami2023SuperresolutionAnalysisMachine}.
The coefficients are hyperparameters, which need to be tuned.
By introducing the governing equations as a penalty term in the loss function, we place a soft constraint on the physics of the prediction. 

Although their excellent capabilities, soft physics constraints may produce trivial solutions \citep{wong2022LearningSinusoidalSpaces}.
Since the importance of the physics depends on the coefficients, the physics loss is ``competing'' with the data loss.
Sometimes networks learn a trivial solution, which satisfies the governing equations without satisfying the data loss, e.g., flows where the velocity is constant everywhere.
Trivial solutions occur because a physics loss of $0$ with a larger data loss gives a better overall loss than a small physics and a small data loss.

The snapshot-enforced loss, proposed by \citet{gao2021SuperresolutionDenoisingFluid}, places a harder constraint on the data loss.
The idea of the strictly enforced data loss is to force the prediction to be equal to the observed values at observed locations so that only the physics loss is minimized during training.
\citet{gao2021SuperresolutionDenoisingFluid} successfully reconstructed steady flows using the snapshot-enforced loss.
However, the snapshot-enforced loss is not robust to noisy observations.

When measurements are taken from experiments, they are often noisy.
Different network structures have been proposed for reconstructing flows from noisy measurements, such as placing restrictions on the shape of the solution \citep{raynaud2022ModalPINNExtensionPhysicsinformed} or using probabilistic networks \citep{maulikromit2020ProbabilisticNeuralNetworks}.
Some information on the flow fields can also be recovered from noisy partial observations by reconstructing the mean instead of the instantaneous flow fields \citep{sun2020PhysicsconstrainedBayesianNeural,sliwinski2023MeanFlowReconstruction}.
\citet{shokar2024StochasticLatentTransformer} in particular, used forced noise and an ensemble network to study the long-term statistics of beta-plane turbulence.

Reconstruction of instantaneous turbulent flows from noisy partial observations has been performed with CNN-based networks in \citep{li2023MultiscaleReconstructionTurbulent,kim2021UnsupervisedDeepLearning,yousif2023DeeplearningApproachReconstructing} as part of a robustness study.
However, these studies require knowledge of the reference data (i.e., the full flow field) during training.

In this paper, we reconstruct the unsteady wake of a triangular cylinder and a turbulent Kolmogorov flow from both clean and noisy sparse measurements to infer the entire flow field.
 Kolmogorov flow and circular cylinder wakes are common test cases in flow reconstruction literature \citep{kelshaw2022PhysicsInformedCNNsSuperResolution,erichson2020ShallowNeuralNetworks,maulikromit2020ProbabilisticNeuralNetworks}, with the wake of a triangular cylinder being important in flame holders in combustion chamber design \citep{agrawal2013ExperimentalStudyFlow,ghani2015LongitudinalTransverseSelfexcited}.
First, we investigate the effect of hard and soft constraints placed on the instantaneous measurements in the loss functions on flow reconstruction with a physics-constrained neural network.
Then, we propose a new loss function for reconstructing flows from noisy measurements.
Finally, we compare the loss functions at different signal-to-noise ratios.

The paper is organized as follows.
First, we introduce two datasets --- a laminar wake and a turbulent Kolmogorov flow in Section~\ref{sec:data}.
Second, we present our methodology, including taking measurements for collocation points, our neural network and different loss functions, in Section~\ref{sec:method}.
Third, we reconstruct the laminar wake from both noisy and non-noisy measurements in Section~\ref{sec:laminar} and reconstruct the Kolmogorov flow from both noisy and non-noisy measurements in Section~\ref{sec:turbulent}.
Finally, we present our conclusion in Section~\ref{sec:conclusion}.

\section{Data and preprocessing}\label{sec:data}

In this study, we reconstruct an unsteady laminar wake of a triangular body and a turbulent Kolmogorov flow. Details on the data generation and preprocessing are given in Section~\ref{sec:data:wake} for the laminar dataset and Section~\ref{sec:data:kol} for the turbulent dataset.

\subsection{Laminar unsteady wake}\label{sec:data:wake}
\begin{figure}[htp]
    \centering
    \includegraphics[width=0.7\textwidth]{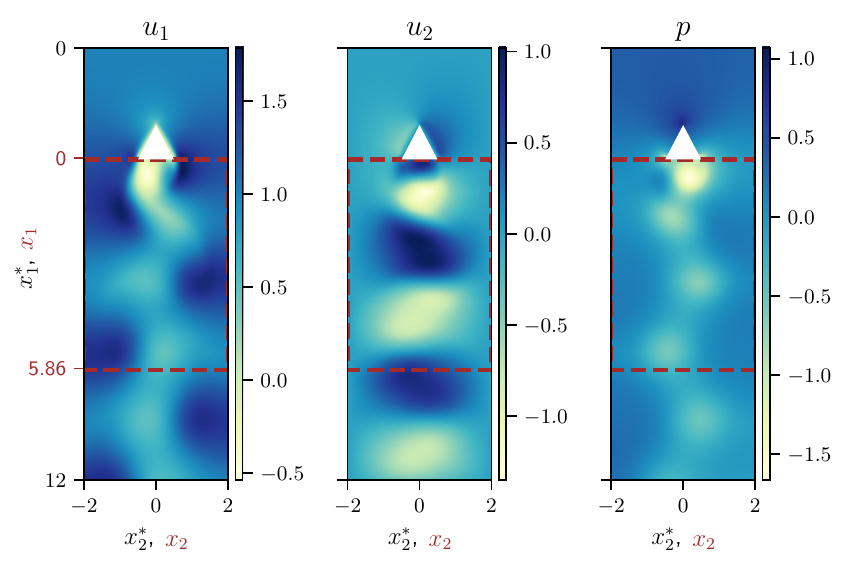}
    \caption{Snapshots of streamwise (left) and wallnormal (centre) velocity and pressure (right) at $t=2.5$, where $t=0$ is the end of the transient period discarded when generating the dataset. The area indicated by the brown box is the area considered by the neural networks, with its coordinates $\vector{x} = [x_1^*-3.0, x_2^*]^T$ shown in brown.}\label{fig:data:domain}
\end{figure}

The incompressible Navier-Stokes equations are cast in a residual fashion for the purpose of this study 
\begin{equation}
    \begin{cases}
        \nabla \cdot \vector{u} = \mathcal{R}_{d} (\vector{u})\\
        \frac{\partial\vector{u}}{\partial t^*} + \vector{u} \cdot \nabla \vector{u} + \nabla p - \frac{1}{Re}\triangle\vector{u} + \vector{f}= \mathcal{R}_{m}(\vector{u},p) ,
    \end{cases}\label{eq:ns-equations}
\end{equation}
where $\vector{u}(\vector{x}^*,t^*) \in \mathbb{R}^{N_u}$ and $p(\vector{x}^*, t^*) \in \mathbb{R}$ are the velocity and pressure at location $\vector{x^*}$ and time $t^*$, $N_u$ is the number of velocity components, and $\vector{f}$ is a forcing term applied to the flow. The unsteady wake of a 2-dimensional triangular body
at Reynolds number $\Re=100$ is generated by direct numerical simulation using Xcompact3D\footnote{Because Xcompact3D only accepts 3D domains, periodic boundary conditions are applied to the spanwise direction to simulate an infinitely-long cylinder and the data presented here is a plane taken from the resulted simulations.}\citep{bartholomew2022Xcompact3dIncompact3d}. The residuals
$\mathcal{R}_d(\vector{u})$ and $\mathcal{R}_m(\vector{u},p)$ are zero when the equations are exactly solved.
For the unsteady wake, $N_u=2$ and $\vector{f}=\vector{0}$.
The computational domain has size $L_1 = 12$ and $L_2 = 4$ in streamwise and wallnormal directions, with 513 and 129 uniformly spaced grid points in each direction, respectively.
The centre of the bottom edge of the equilateral triangle is placed at (3.0,2.0) oriented such that the base is perpendicular to the streamwise velocity.
Lengths are nondimensionalized by the length of the side of the equilateral triangle.
Zero gradients are applied at $x_1^* = 12$, and slip walls are applied at $x_2^* = -2$ and $x_2^* = 2$;
The time step is $\Delta t^* = 0.0002$.
Equation~\eqref{eq:ns-equations} are solved using the 3\textsuperscript{rd}-order Adams-Bashforth scheme in time and the 6\textsuperscript{th}-order compact scheme in space \citep{laizet2009HighorderCompactSchemes}.
When solved numerically, the simulated flow has near zero residuals, $[\mathcal{R}_d, \mathcal{R}_m]^T \approx \vector{0}$.
The transient period from $t^*=0$ to $100$ is discarded to ensure that the final dataset consists of only periodic vortex shedding \citep{gangaprasath2014EffectsAspectRatio} at a statistically stationary regime.
From here on, $t=0$ denotes the time of the first snapshot of the final dataset, i.e. $t = t^*-100$.
A snapshot of the flow is saved every 625 steps, resulting in a time step $\Delta t = 0.125$ for the final dataset, i.e.\ over 40 snapshots per vortex shedding period.
Figure~\ref{fig:data:domain} shows the data at $t=2.5$ on the plane $x_3^*=0$. 
Only data in the brown box of figure~\ref{fig:data:domain} is considered by the neural networks for computational efficiency, capturing the near wake and the parallel vortex shedding.
For constructing the final dataset, a snapshot of a flow (area within the brown box) contains the grid points on the plane at $x_3^*=0$ bounded by $x_1^*=3.0$, $x_1^*=8.86$, $x_2^*=-2$ and $x_2^*=2$.
For simplicity, we define the coordinates of the dataset to be $\vector{x} = {[x_1^*-3.0, x_2^*]}^T$, shown in figure~\ref{fig:data:domain} in brown.

The snapshots are organized into snapshot matrices $\mat{U} \in \mathbb{R}^{N_t \times N_1 \times N_2 \times N_u}$ and $\mat{P} \in \mathbb{R}^{N_t\times N_1 \times N_2 \times 1}$ for velocity and pressure, where $N_t$ is the number of snapshots, $N_1 = 250$ is the number of grid points in the $x_1$ direction and $N_2 = 129$ is the number of grid points in the wallnormal direction.
The dataset $\mat{D}^T = [\mat{U}^T, \mat{P}^T]$ is the snapshot matrix of the full state.
Each full state vector is  $\mat{D}(\vector{x},t)^T = [\vector{u}^T,p]$, where $\vector{u}^T = [u_1(\vector{x},t),u_2(\vector{x},t)]$ is the 2D velocity vector and $p = p(\vector{x},t)$ is the pressure.
The laminar dataset contains 600 snapshots, $N_t=600$, which cover approximately 15 vortex shedding periods.

\subsection{Turbulent Kolmogorov flow}\label{sec:data:kol}

\begin{figure}[htb]
    \centering
    \includegraphics[width=0.9\linewidth]{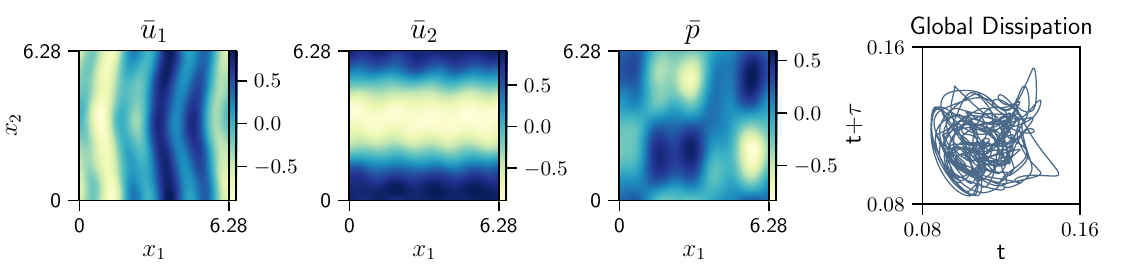}
    \caption{The time-averaged velocity in $x_1$ direction ($u_1$), $x_2$ direction ($u_2$), pressure ($p$) and global dissipation with a time delay of $\tau\approx5$ (left to right).}\label{fig:data:kol}
\end{figure}

KolSol \citep{kelshawdanielMagriLabKolSolPseudospectral}, a pseudo-spectral solver, is used to generate Kolmogorov flows by solving~\eqref{eq:ns-equations} with the forcing term $\vector{f} = \vector{e}_1\sin{k\vector{x}}$, where $\vector{e}_1$ is the unit vector ${[1,0]}^T$, in a 2D periodic box with length $2\pi$ on both sides. 
The Kolmogorov flows are generated at $\Re=34$, with 32 wavenumbers, the forcing frequency $k=4$ and the timestep $\triangle t^* = 0.01$, resulting in weakly turbulent flows \citep{racca2023NeuralNetworksPrediction}.
The flow in the frequency domain is then converted into the physical domain on a 128-by-128 grid. 
A snapshot is saved every 10 $\triangle t^*$, resulting in $\triangle t = 0.1$ for the datasets.
The end of the transient period is $t=0$.
Two datasets are generated by using different initial random seeds --- one for the grid sensitivity study in Section~\ref{sec:appendix:2dkol-grid-sensitivity} and one for the results in Section~\ref{sec:turbulent:clean} and~\ref{sec:turbulent:noisy}. 
Each Kolmogorov dataset, $\mat{D} = {[\mat{U}^T, \mat{P}^T]}^T \in \mathbb{R}^{N_t \times N \times N \times (N_u+1)}$, is the snapshot matrix of the full state with velocity and pressure, where  $N_t=6000$, $N=128$ and $N_u=2$.
Figure~\ref{fig:data:kol} shows an overview of a Kolmogorov flow dataset.
The global dissipation is the local dissipation $\| \nabla\vector{u} \|^2/\Re$ averaged over the domain.

\section{Methodology}\label{sec:method}
In this section, we first present how the measurements used in the reconstruction are selected from our simulated datasets (Section~\ref{sec:method:observations}).
We then present the physics-constrained dual-branch convolutional neural network (Section~\ref{sec:method:network}).
Finally, we present the different formulations of loss functions we will investigate (Section~\ref{sec:method:loss}).

\subsection{Taking measurements}\label{sec:method:observations}
From each dataset, we select a set of sparse measurements from which we will reconstruct the flows.
This set of sparse measurements is the only data-related information available for the network in training, i.e., no high-resolution data is needed in the methods that we propose.
Measurements of all velocity components and pressure are taken at sensor locations $\vector{x}_{s}$ (general sensors).
Additional pressure measurements are taken at input locations $\vector{x}_{in}$.
The separate input pressure measurements simulate a common situation in bluff body wake experiments, e.g.~\citep{rowand.brackston2017FeedbackControlThreeDimensional}.
For the bluff body wake, $\vector{x}_{in}$ is selected to be immediately behind the triangular body; 
for the Kolmogorov flow, $\vector{x}_{in}$ is randomly selected.
The set of measurements $\xi\left( \mat{D} \right)$ is a collection of the input pressure, and the velocities and pressure measured at the sensor locations, $\xi\left( \mat{D} \right) = \left\{ \mat{U}(\mat{x}_s), \mat{P}(\mat{x}_s), \mat{P}(\mat{x}_{in}) \right\}$.
The measurements are taken at different locations for different tests. The exact locations will be shown together with the test results in Section~\ref{sec:laminar} and Section~\ref{sec:turbulent}.

\subsection{Physics-constrained dual-branch convolutional neural network}\label{sec:method:network}

We develop a physics-constrained dual-branch convolutional neural network (PC-DualConvNet) for the reconstructions, inspired by \citep{kelshaw2022PhysicsInformedCNNsSuperResolution,shokar2024StochasticLatentTransformer,ronneberger2015UNetConvolutionalNetworks,li2021FourierNeuralOperator}.
The advantage of the PC-DualConvNet is that it does not require the full state to be known at all grid points during training, and that the measurements do not need to be in a regular grid.
The goal of the network is to infer unseen spatial information for a given set of measurements.

\begin{figure}[htb]
    \centering
    \includegraphics[width=0.9\linewidth]{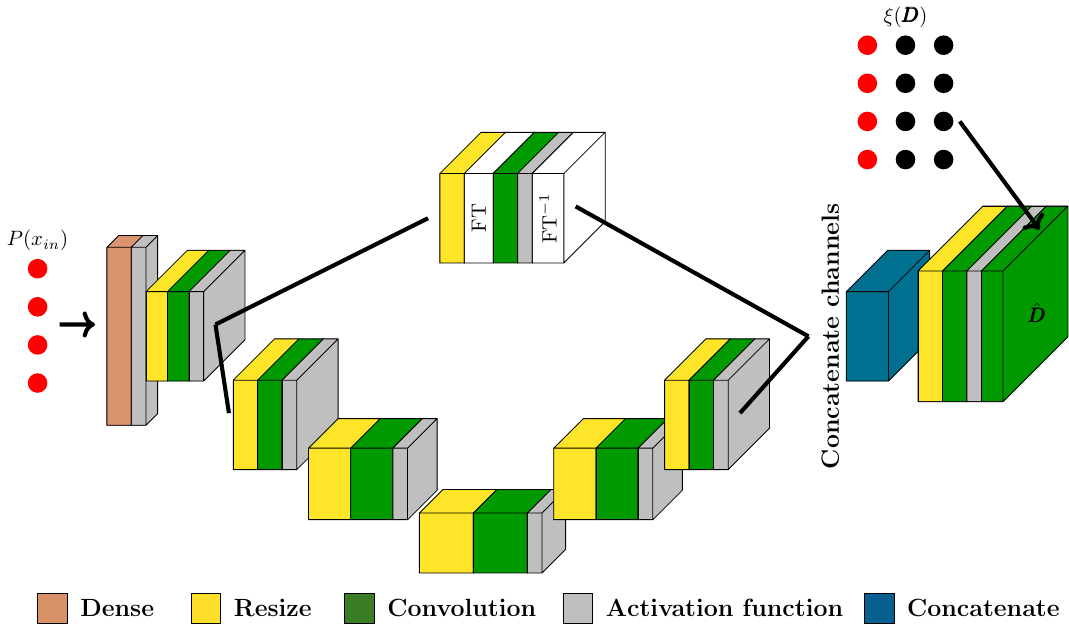}
    \caption{Schematic of the two-branch Fourier convolutional neural network. The number and size of layers  are for illustration. The Fourier Transform layer ($\sym{FT}$) and the inverse Fourier Transform layer ($\sym{FT}^{-1}$) are optional and are activated (or deactivated)  for different loss functions.}\label{fig:method:network}
\end{figure}

PC-DualConvNet aims to produce the flow field $\hat{\mat{D}}$ to approximate the data $\mat{D}$ from pressure input $\mat{P}_{in}$, using sparse measurements $\xi(\mat{D})$ as collocation points.
PC-DualConvNet consists of a U-Net style \citep{ronneberger2015UNetConvolutionalNetworks} convolutional branch, which is common in image processing (lower branch in Figure~\ref{fig:method:network}), and an optional Fourier branch for reducing the spatial aliasing of the output (top branch in Figure~\ref{fig:method:network}).
The Fourier Transform is performed over both temporal and spatial dimensions of the dataset.
During training, we find that the Fourier Transform in the upper branch is particularly helpful in improving the robustness of the snapshot-enforced loss $\lossStrict$ (Section~\ref{sec:method:loss}).
The benefit of activating the Fourier layer varies with different datasets and is considered as a hyperparameter.
Depending on the test case, as explained in the later sections, the top branch may perform Fourier Transform and inverse Fourier Transform immediately before and after the convolution layers.
The output of the top and lower branches are stacked channel-wise before the final convolution layers, which produce the output $\hat{\mat{D}}$.

The PC-DualConvNet's parameters $\mat{w}$ are the (local) minimizers of a loss function $\loss$, which measures the error between $\hat{\mat{D}}$ and $\mat{D}$. 
The optimization process is 
\begin{equation}
    \mat{w}^* = \argmin_{\mat{w}} \loss(\mat{P}(\vector{x}_{in});\mat{w}),
    \label{eq:optimisation}
\end{equation}
in which the design of the loss formulation $\loss$ is a goal of this paper and will be discussed in detail in Section~\ref{sec:method:loss}.
The  reconstruction error is  the relative $\ell_2$ error
\begin{equation}
    \epsilon (\%) = \sqrt{\frac{\| \hat{\mat{D}} - \mat{D} \|^2_2}{\| \mat{D} \|^2_2}}.\label{eq:method:rel-l2}
\end{equation}

\subsection{Loss functions}\label{sec:method:loss}
The loss $\loss$ is the quantity to be minimized when training a neural network (Equation~\eqref{eq:optimisation}).
The governing equations are incorporated into the loss formulations for the physical reconstruction.
We define the momentum and continuity loss terms of the dataset $\mat{D}$ to be the $\ell_2$ norm of the residuals of the continuity and momentum equations, as defined in Equation~\eqref{eq:ns-equations} at all grid points, as
\begin{gather}
    \loss_{div}(\mat{D}) = \| \mathcal{R}_d(\mat{U}) \|^2_2, \\
    \loss_{mom}(\mat{D}) = \| \mathcal{R}_m(\mat{U,P}) \|^2_2.
    \label{eq:method:residual-loss}
\end{gather}
The derivatives are computed with second-order finite difference schemes on the boundaries, and fourth-order schemes at the interior points.
A data-related loss is also needed for the output to minimize the error with the measurements selected from the reference datasets.
We define the sensor loss to be the $\ell_2$ norm of the difference of the output $\hat{\mat{D}}$ and the reference data $\mat{D}$ at observed locations, as 
\begin{equation}
    \loss_o (\hat{\mat{D}}, \mat{D}) = \| \xi(\hat{\mat{D}}) - \xi(\mat{D}) \|^2_2.
    \label{eq:method:sensor-loss}
\end{equation}

\subsubsection{Softly-constrained loss}\label{sec:loss:pinn}

A widely applied method to include physics into the network is to define the loss function to be the weighted sum of a sensor loss and a physics loss, applied in various superresolution convolutional neural networks and physics-informed neural networks \citep{raissi2020HiddenFluidMechanics,kelshaw2022PhysicsInformedCNNsSuperResolution,raissi2019PhysicsinformedNeuralNetworks,fukami2023SuperresolutionAnalysisMachine}.
We define the softly-constrained loss $\lossClassic$ to be
\begin{equation}
    \lossClassic(\mat{P}_{in};\mat{w}) = \lambda_o \loss_o(\hat{\mat{D}}, \mat{D}) + \lambda_{div} \loss_{div}(\hat{\mat{D}}) + \lambda_{mom} \loss_{mom}(\hat{\mat{D}}),
    \label{eq:pinn-loss}
\end{equation}
where $\lambda$ denotes the non-negative regularisation factor. 
Both the physics of the output and the values of the output at the sensor locations are included in the loss function as regularisation terms, which means $\loss_o$ and $\loss_p$ will be minimized, but not guaranteed to be a specific value, hence its name `softly-constrained'.

The components of the softly-constrained loss can be grouped into the sensor loss $\loss_o$ and the physics loss $\loss_p$ which is the combined continuity and momentum loss $\loss_{div} + \loss_{mom}$.
A trivial solution is a solution which has $\loss_p = 0$ but a large $\loss_o$, meaning that the network finds another solution to the governing equation that is different from the reference datasets.
An example of a trivial solution is a flow field that is constant everywhere.
Trivial solutions may develop when both $\loss_p$ and $\loss_o$ cannot, theoretically, be $0$ at the same time, which can be caused by noise, either experimental or numerical.
When reconstructing a synthetic dataset, the source of the noise could come from the use of different numerical schemes when estimating the derivatives. 
For example, the residuals in $\loss_p$ for the reconstructed laminar dataset, defined in~\eqref{eq:method:residual-loss}, are computed with a mix of second- and fourth-order finite differences.
However, the laminar dataset is generated with a sixth-order scheme (see Section~\ref{sec:data:wake}).
Therefore, if the measurements at sensor locations from the reference and the reconstructed datasets are the same, meaning $\loss_o(\hat{\mat{D}}, \mat{D})=0$, and the reference dataset has $\loss_p(\hat{D})$ when using the sixth-order scheme, then $\loss_p(\hat{\mat{D}})$ cannot be zero when using the mix of second- and fourth-order scheme.
The possible scenarios are summarized in Table~\ref{tab:pinn-loss-trivial-solution-scenarios}.
Scenario~1 is not possible when different numerical schemes are used in data generation and training, or when the reference data is collected from experiments. 
Depending on the sensor placements, scenario~2 may have a theoretical sensor loss close to $0$ (when using a few sensors) or much larger than $0$ (large numbers of sensors or good sensor locations).
Scenarios~3 and~4 are acceptable reconstructions, but their theoretical minimum total losses are also not $0$.
Therefore, the network may produce trivial solutions instead of the correct reconstructed flows.
In other words, given that the ideal scenario~1 is not possible, we want to be in scenarios~3 or~4, but they have the same non-zero theoretical minimum total loss as the unwanted scenario~2.
\begin{table}[htb]
    \caption{Possible solutions and their corresponding theoretical loss values when training with the softly-constrained loss. The network needs to minimize both physics loss and sensor loss.
    Scenario 1 has a theoretical minimum loss of $0$ and is ideal, but is also unlikely when different numerical schemes are used in data generation and loss computation, or if data is experimental.
    Scenario 2 is the trivial solution, where the outputs satisfy the governing equations but not the measurements. Depending on the sensor placements, the minimum $\loss_o$ could be either approximately $0$ or much larger than $0$.
    Scenarios~3 and~4 are desired, but they also have a theoretical minimum total loss larger than $0$.
    }\label{tab:pinn-loss-trivial-solution-scenarios}
    \begin{tabular}{|c|c|c|c|c|c|}
        \hline
         & Total loss value & Sensor loss & Physics loss & Numerical scheme & Reconstructed flow \\ 
        \hline
        1 & 0 & 0 & 0 & same & $\hat{\mat{D}} = $\mat{D} \\
        2 & $>$0 or $\approx$0 & $>$0 or $\approx$0 & 0 & different & trivial solution \\
        3 & $\approx$0 & 0 & $\approx$0 & different & $\hat{\mat{D}} = $\mat{D} \\
        4 & $\approx$0 & $\approx$0 & $\approx$0 & different & $\hat{\mat{D}} \approx $\mat{D} \\
        \hline
    \end{tabular}
\end{table}

\subsubsection{Snapshot-enforced loss}\label{sec:loss:strict}
To avoid trivial solutions and improve the training, we strictly enforced the measurements and minimize physics loss.
The snapshot-enforced loss for steady flows was proposed by \citep{gao2021SuperresolutionDenoisingFluid}.
The snapshot-enforced loss minimizes only the physics loss while the sensor loss is enforced to be~$0$, providing a harder constraint on the predicted flow at sensor locations compared to the softly-constrained
By placing a harder constraint on the measurements, we also eliminate trivial solutions.
In this paper, we define the snapshot-enforced loss $\lossStrict$ as
\begin{equation}
    \lossStrict = \lambda_{div} \loss_{div}(\mat{\Phi}) + \lambda_{mom} \loss_{mom}(\mat{\Phi}),
    \label{eq:strictly-enforced-loss}
\end{equation}
where $\mat{\Phi}^T = [\mat{\Phi}_u^T, \mat{\Phi}_p^T]$ is defined as 

\begin{minipage}{0.5\linewidth} 
    \begin{align*}
        \mat{\Phi_u}(\vector{x}) =
        \begin{cases}
            \mat{U}(\vector{x}) & \text{where } \vector{x} \in \mat{x}_s, \\
            \hat{\mat{U}}(\vector{x}) & \text{otherwise}.
        \end{cases}
    \end{align*}
\end{minipage}%
\begin{minipage}{0.5\linewidth} 
    \begin{align}
        \mat{\Phi_p}(\vector{x}) =
        \begin{cases}
            \mat{P}(\vector{x}) & \text{where } \vector{x} \in \{\mat{x}_s, \mat{x}_{in}\}, \\
            \hat{\mat{P}}(\vector{x}) & \text{otherwise}.
        \end{cases}
    \end{align}
\end{minipage}

\begin{figure}[htb]
    \centering
    \includegraphics[width=0.9\linewidth]{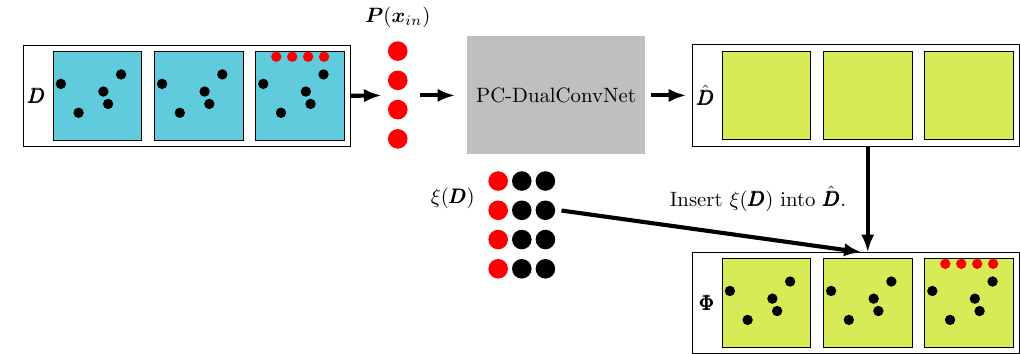}
    \caption{Computing the snapshot-enforced loss.}\label{fig:loss:strict}
\end{figure}

Figure~\ref{fig:loss:strict} is a graphical representation of training a PC-DualConvNet with the snapshot-enforced loss $\lossStrict$.

\subsubsection{Mean-enforced loss}\label{sec:loss:mean}
The snapshot-enforced loss $\lossStrict$ ensures the reconstructed flow fits the measurements, but it also enforces the noise on the measurements if the measurements are noisy.
\citep{gao2021SuperresolutionDenoisingFluid} suggest using the softly-constrained loss for noisy measurements.
In this paper, to deal with noisy measurements, we propose the mean-enforced loss.
The mean-enforced loss $\lossMean$ combines the advantages of the softly-constrained and the snapshot-enforced loss, defined as
\begin{equation}
    \lossMean = \lambda_o \loss_o(\hat{\mat{D}}, \mat{D}) + \lambda_{div} \loss_{div}(\mat{\Phi}) + \lambda_{mom} \loss_{mom}(\mat{\Phi}),
    \label{eq:mean-loss}
\end{equation}
where $\mat{\Phi}^T = [\mat{\Phi}_u^T, \mat{\Phi}_p^T]$ is 

\begin{minipage}{2.4in} 
    \begin{align*}
        \mat{\Phi_u}(\vector{x}) =
        \begin{cases}
            \overline{\mat{U}}(\vector{x}) + \hat{\mat{U}}^\prime(\vector{x}) \; \text{where} \; \vector{x} \in \mat{x}_s, \\
            \hat{\mat{U}} \; \text{otherwise}.
        \end{cases}
    \end{align*}
\end{minipage}%
\begin{minipage}{4in} 
    \begin{align}
        \mat{\Phi_p}(\vector{x}) =
        \begin{cases}
            \overline{\mat{P}}(\vector{x}) + \hat{\mat{P}}^\prime(\vector{x}) \; \text{where} \; \vector{x} \in \{ \mat{x}_s, \mat{x}_{in} \}, \\
            \hat{\mat{P}} \; \text{otherwise}.
        \end{cases}
    \end{align}
\end{minipage}

\begin{figure}[htb]
    \centering
    \includegraphics[width=0.9\linewidth]{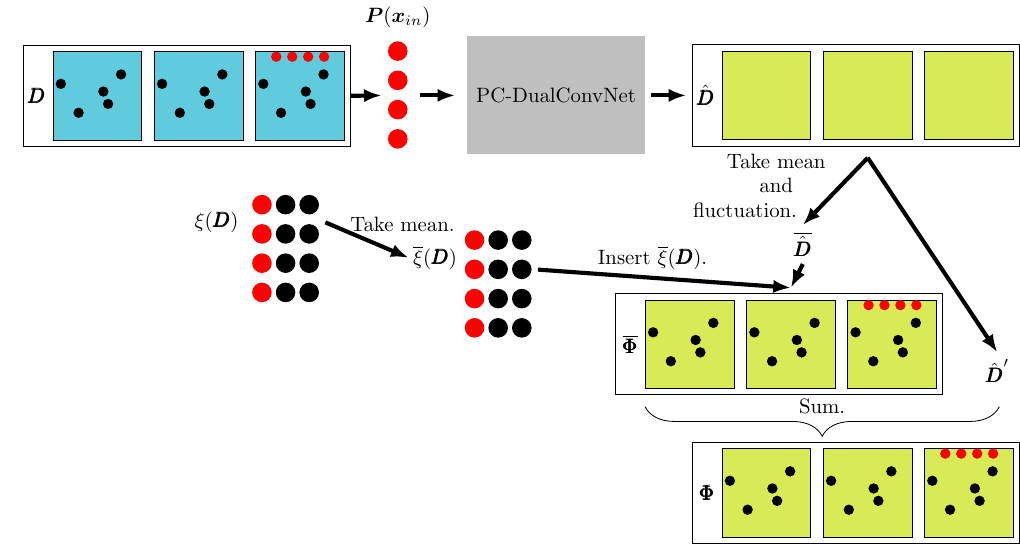}
    \caption{Computing the mean-enforced loss. The mean is taken over a batch of inputs during training.}\label{fig:loss:mean}
\end{figure}

The symbols $\overline{*}$ and $*^\prime$ denote the time-averaged and fluctuating components of any variable~$*$, respectively.
Figure~\ref{fig:loss:mean} is a graphical representation of $\lossMean$.
By changing the ratios of the coefficients $\lambda$, we can adjust how much smoothing to apply when reconstructing from noisy observation.
The smaller the $\lambda_o$ compared to $\lambda_{div}$ and $\lambda_{mom}$, the stronger the smoothing.
\FloatBarrier%

\section{Reconstructing the laminar wake}\label{sec:laminar}
In this section, we reconstruct the laminar wake from both non-noisy measurements and noisy measurements.
We first reconstruct the laminar wake from 18 sparse sensors placed in the domain in Section~\ref{sec:laminar:clean}.
We then compare the three loss functions introduced in Section~\ref{sec:method:loss} in reconstructing the wake from sparse measurements contaminated with white noise.

\subsection{Reconstructing the laminar wake from non-noisy measurements}\label{sec:laminar:clean}
\begin{figure}[htb]
    \centering
    \includegraphics[width=3in]{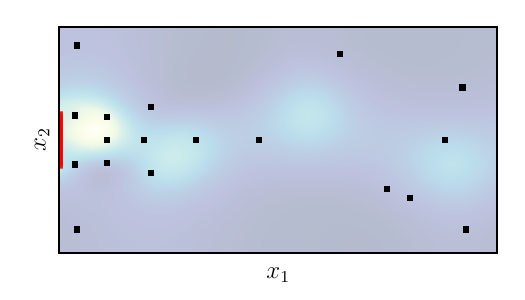}
    \caption{Sensor placement for laminar wake, non-noisy measurements. The measurements consists of the pressure inlet measurement at $\vector{x}_{in}$ (red) and measurements of all components at $\vector{x}_s$ (black).}\label{fig:2dtriangle:clean-sensor-location}
\end{figure}

For all reconstructions of the laminar wake, pressure sensors are placed on all grid points immediately behind the bluff body (31 in total) at $\vector{x}_{in}$.
These pressure measurements $\mat{P}(\vector{x}_{in})$ are used as the input to the network (red dots in Figure~\ref{fig:2dtriangle:clean-sensor-location}).
When reconstructing from non-noisy measurements, 12 general sensors (black dots in Figure~\ref{fig:2dtriangle:clean-sensor-location}) are placed at the maximum and minimum of the first two leading POD modes of the pressure and velocity components \citep{cohen2003SensorPlacementBased}. 
A further 6 general sensors are placed randomly near the edge of the domain. 
The sensor locations $\vector{x}_s$ account for only $\sim0.06\%$ of all grid points.

The softly-constrained loss $\lossClassic$ and the snapshot-enforced loss $\lossStrict$ are tested for their ability to reconstruct the wake. 
For each loss function, a hyperparameter optimization is carried out\footnote{Using Bayesian hyperparameter optimization provided by Weights and Biases (http://wandb.ai/).} and the hyperparameters that minimize total loss $\loss_p + \loss_o$ are selected.
(There is a significant difference in the reconstruction results between $\lossClassic$ and $\lossStrict$ when applied to reconstruct the laminar wake from non-noisy measurements.)

\begin{figure}[hbt]
    \centering
    \begin{subfigure}{0.8\linewidth}
        \includegraphics[width=\linewidth]{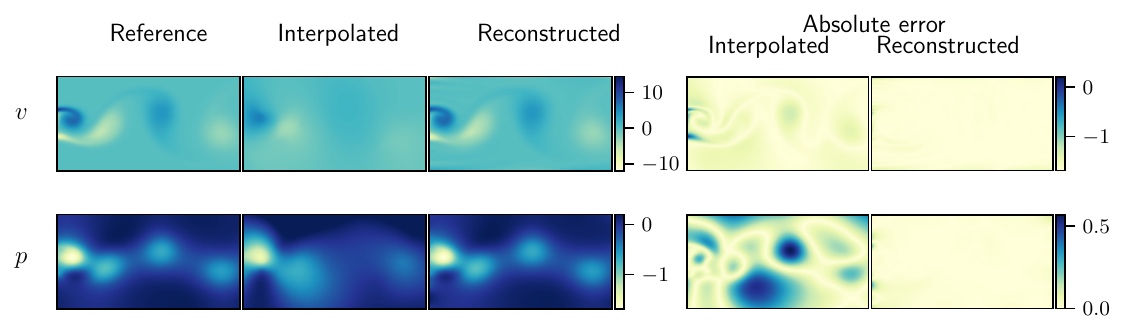}
        \caption{}\label{fig:2dtriangle:clean-snapshots}
    \end{subfigure}
    \begin{subfigure}{0.8\linewidth}
    \includegraphics[width=\linewidth]{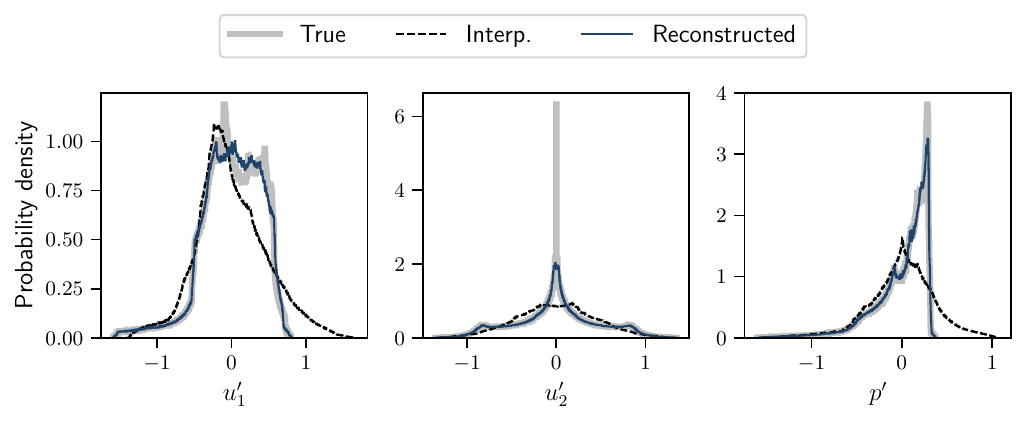}
    \caption{}\label{fig:2dtriangle:clean-probability-density}
    \end{subfigure}
    \caption{Reconstructed laminar wake from non-noisy measurements.
        (a) Snapshots of the reconstructed laminar wake compared with the reference and interpolated snapshot. The last two columns show the instantaneous absolute error of the interpolated snapshot and the reconstructed snapshot, respectively.
        (b) The probability density of the fluctuating components of the entire flow.
        Both (a) and (b) show that the reconstructed flow is a closer match with the reference data than mere interpolation.
        }\label{fig:2dtriangle:clean}
\end{figure}

The model with the lowest total loss in this case uses $\lossClassic$ and is used to generate the results in this section (see Table~\ref{tab:ap:laminar:clean-params} for the network structures and hyperparameters).
A snapshot of the reconstructed flow is compared with the reference and Interpolated snapshot in Figure~\ref{fig:2dtriangle:clean-snapshots}, which shows that the reconstructed flow resembles closely to the reference data and has a smaller error.
Vorticity $\vector{v}$ is the cross product of the velocity $\nabla\times\vector{u}$.
The probability density of the fluctuating components of the reconstructed flow also matches more closely to the reference data than the interpolated (Figure~\ref{fig:2dtriangle:clean-probability-density}).
\begin{table}[htb]
    \centering
    \caption{Summary of the losses for the reconstructed laminar wake from non-noisy measurements. Values are shown in the format \textit{mean} $\pm$ \textit{standard deviation} calculated over 5 runs initiated with different random weights.}\label{tab:2dtriangle:clean}
    \begin{tabular}{|c|c|c|}
        \hline
         & \textbf{$\epsilon$ (\%)} & \textbf{$\loss_p$} \\
         \hline
        Reference data & N/A & 0.04338$\pm$0.00000 \\
        Interpolated & 43.26$\pm$0.00 & 1.06721$\pm$0.00000 \\
        Reconstructed & 3.34$\pm$0.18 & 0.00028$\pm$0.00002 \\
        \hline
    \end{tabular}
\end{table}
Table~\ref{tab:2dtriangle:clean} compares the relative error and the physics loss of the reconstructed flow with the reference data and the interpolated flow.
Interpolation of the laminar wake is performed per snapshot using thin-plate spline to handle the very sparse and irregular data points.
The physics loss of the reference data is non-zero because the physics loss in training is computed with a mix of second- and fourth-order numerical schemes when a different numerical scheme is used in data generation.
Detailed explanations on the effect of using different numerical schemes are given in Section~\ref{sec:data:wake} and~\ref{sec:method:loss}.
The small errors numerically show that the network has successfully reconstructed the laminar wake from sparse measurements.

\subsection{Reconstructing the laminar wake from noisy observations}\label{sec:laminar:noisy}
\begin{figure}[htb]
    \centering
    \includegraphics[width=3in]{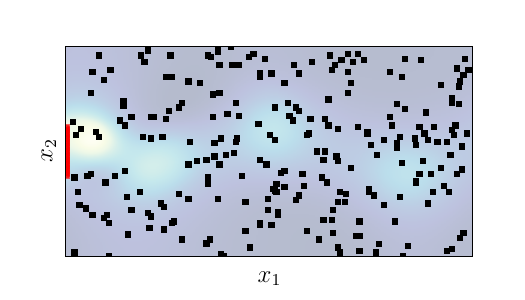}
    \caption{Sensor placement for laminar wake, noisy measurements. The measurements consist of the pressure inlet measurement at $\vector{x}_{in}$ (red) and measurements of all components at $\vector{x}_s$ (black).}\label{fig:2dtriangle:noisy-sensor-location}
\end{figure}

For reconstructing the laminar wake from noisy measurements, 250 general sensors are randomly placed in the domain (Figure~\ref{fig:2dtriangle:noisy-sensor-location}). 
The sensor locations $\vector{x}_s$ account for only $\sim 0.8\%$ of all grid points.
The noise $\vector{e} \in \mathbb{R}^{N_u+1}$ for the full state $\mat{D}(\vector{x},t)$ at each discrete grid point is generated from a Gaussian distribution 
$\vector{e} \sim \mathcal{N}(\vector{0},\vector{\sigma_e})$,
where the components of $\vector{\sigma}_e$ correspond to the components of the flows $[\vector{u},p]$.
The level of noise is measured by the signal-to-noise ratio (SNR), defined as
$\SNR = 10 \log \left( \sigma^2_{i} / \sigma^2_{e_i} \right),$
where $\sigma_{i}$ and $\sigma_{e_i}$ denotes the standard deviation of the $i$-th component of the flows.
The noisy data is $\mat{D}_n(\vector{x},t) = \mat{D}_n(\vector{x},t) + \vector{e}$ ($\vector{e}$ is different at each discrete grid point);
the noisy measurements are $\xi(\mat{D}_n)$;
the inputs to the network is $\mat{P}_n(\vector{x}_{in})$.
Three different $\SNR$s are considered in this paper, $\SNR=20,10$ and $5$.

\begin{figure}[htb]
    \centering
    \includegraphics[width=0.8\linewidth]{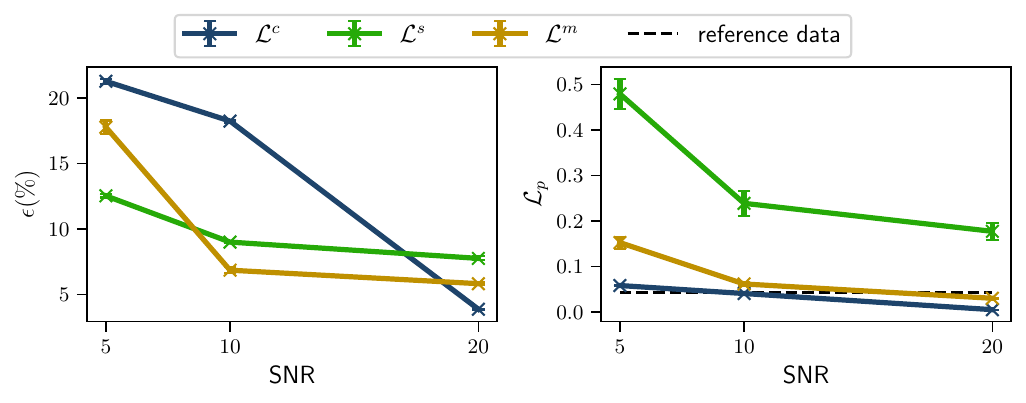}
    \caption{Comparison of loss functions for different SNR.\@ The reference physics loss is the residual of the Navier-Stokes equations solved numerically. The mean (markers) and standard deviation (error bars) are calculated over five repeats with randomly initialized weights.}\label{fig:2dtriangle:noisy-compare-lossfn}
\end{figure}
For consistency of the results, the same random seed is used to generate the noise for all test cases shown in this section.
The hyperparameters for the networks are shown in Table~\ref{tab:ap:laminar:noisy-params-classic} to~\ref{tab:ap:laminar:noisy-params-mean}.
Figure~\ref{fig:2dtriangle:noisy-compare-lossfn} summarises the effect of different loss functions on reconstructing the laminar wake from noisy, sparse measurements.
At $\SNR=20$, all loss functions result in reconstruction errors under 10\% (Figure~\ref{fig:2dtriangle:noisy-compare-lossfn} left panel), showing that all loss functions can lead to good reconstruction with small levels of noise.
The reconstruction error of the softly-constrained loss reaches above 10\% by $\SNR=10$ whilst the errors of the strictly enforced and the mean-enforced loss only reach above 10\% at $\SNR=5$.
The slower rate of increase of error as the measurements become noisier shows that losses with harder constraints are less sensitive to noise.

Figure~\ref{fig:2dtriangle:noisy-compare-lossfn} (right panel) shows a comparison of the physics loss $\loss_p$ for the networks using the different loss functions.
At any level of noise, $\lossStrict$ leads to the highest $\loss_p$ amongst the three loss functions tested.
$\lossMean$ also has a higher physics loss than $\lossClassic$, but  close to the reference physics loss.
The harder constraint on the instantaneous measurements in $\lossStrict$ does not allow for de-noising, thus leading to a higher $\loss_p$.
Since $\lossMean$ is specifically designed to avoid placing hard constraints on the instantaneous measurements, the impact of noise on the reconstruction error of $\lossMean$ is smaller compared to $\lossStrict$.
\begin{figure}
    \centering
    \begin{subfigure}{\linewidth}
        \includegraphics[width=0.9\linewidth]{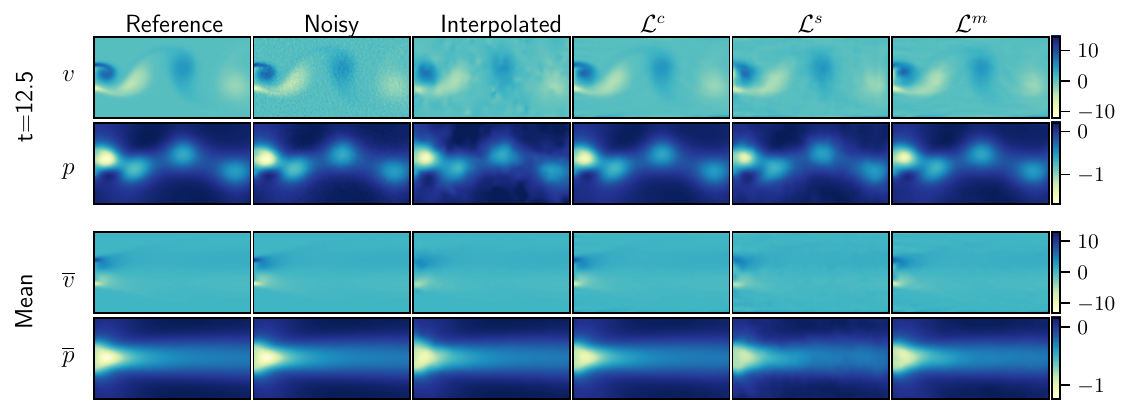}
        \caption{}\label{fig:2dtriangle:noisy-snapshots20}
    \end{subfigure}
    \begin{subfigure}{\linewidth}
        \includegraphics[width=0.9\linewidth]{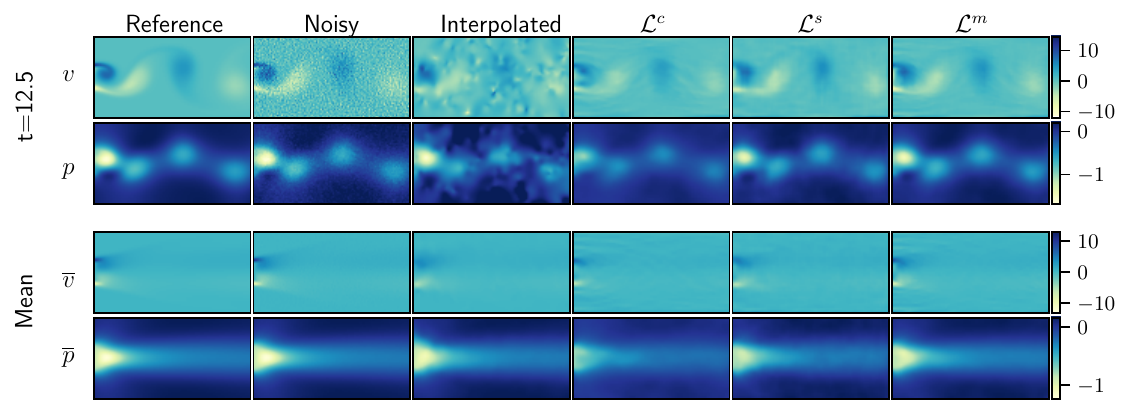}
        \caption{}\label{fig:2dtriangle:noisy-snapshots10}
    \end{subfigure}
    \begin{subfigure}{\linewidth}
        \includegraphics[width=0.9\linewidth]{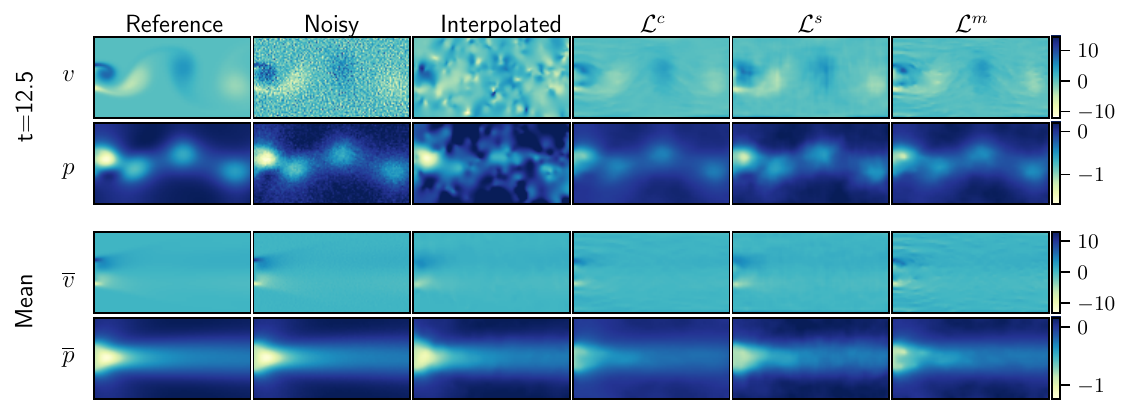}
        \caption{}\label{fig:2dtriangle:noisy-snapshots5}
    \end{subfigure}
    \caption{Reconstructed laminar wake from noisy measurements at SNR=20 (\ref{fig:2dtriangle:noisy-snapshots20}), 10 (\ref{fig:2dtriangle:noisy-snapshots10}) and 5 (\ref{fig:2dtriangle:noisy-snapshots5}).}\label{fig:2dtriangle:noisy-snapshots}
\end{figure}
The effect of noise can be appreciated in Figure~\ref{fig:2dtriangle:noisy-snapshots}, in which snapshots of the instantaneous and the mean reconstructed wake are compared with the reference and the interpolated.
At $\SNR=10$ (Figure~\ref{fig:2dtriangle:noisy-snapshots10}), the wake reconstructed with $\lossClassic$ shows the same flow features as the reference, but with a reduced variation of the flow field (the difference between the maximum and minimum value is smaller than the reference).
A reduced variation of the predicted flow field is the first sign of a network breaking down.
In comparison, the wake reconstructed with $\lossMean$ and $\lossStrict$ at $\SNR=10$ and at $\SNR=20$ shows no visible difference.
At both $\SNR$s, the wake reconstructed with $\lossStrict$ shows visible noise in the snapshots, which is not present in the snapshots reconstructed with $\lossMean$.
$\lossMean$ and $\lossStrict$ also start to show a reduced variation of the flow field at $\SNR=5$, even though the reduction is stronger for $\lossMean$.
At $\SNR=5$, the mean of the interpolated data is more similar to the reference than the reconstructed, particularly in the near wake.
However, at all SNRs, the instantaneous snapshots reconstructed wake resemble the reference more closely than the interpolated, showing that regardless of noise, the PC-DualConvNet can reconstruct the flow with a lower relative error than interpolation.
Both the quantitative (Figure~\ref{fig:2dtriangle:noisy-compare-lossfn}) and qualitative comparison of the loss functions (\ref{fig:2dtriangle:noisy-snapshots}) show that applying a harder constraint on the measurements delays the breakdown of the networks, so the networks are more robust to noise.
At $\SNR=20$ and~$10$, $\lossMean$ is able to achieve the same level of de-nosing as $\lossClassic$ while being more robust to noise, making it more suitable for medium levels of noise in the measurements.

\section{Reconstructing the turbulent Kolmogorov flow}\label{sec:turbulent}
In this section, we apply and compare the three loss functions described in Section~\ref{sec:method:loss} to reconstruct 2D turbulent Kolmogorov flows from sparse measurement.

\subsection{Reconstructing from non-noisy measurements}\label{sec:turbulent:clean}

\begin{figure}[htb]
    \centering
    \includegraphics[width=1.7in]{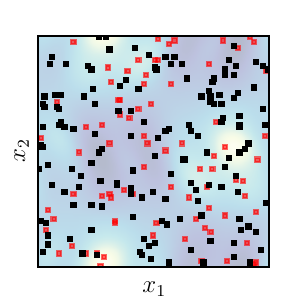}
    \caption{Random sensors used to reconstruct turbulent Kolmogorov flow from non-noisy measurements, showing the input sensors (red) and general sensors (black).}\label{fig:2dkol:clean_sensor_locations}
\end{figure}

In this section, we reconstruct the Kolmogorov flow from 80 input sensors and 150 general sensors placed randomly in the domain.
The general sensors account for approximately 0.9\% of all grid points.
The sensor placement is shown in Figure~\ref{fig:2dkol:clean_sensor_locations}.
The number of input and general sensors are determined through a grid sensitivity study, which has been included in Appendix~\ref{sec:appendix:2dkol-grid-sensitivity}. 
The network parameters for both the current section (Section~\ref{sec:turbulent:clean}) and the next section (Section~\ref{sec:turbulent:noisy}) are
found in Table~\ref{tab:ap:2dkol:params-classic} to~\ref{tab:ap:2dkol:params-mean}.

\begin{figure}[htb]
    \centering
    \includegraphics[width=5.5in]{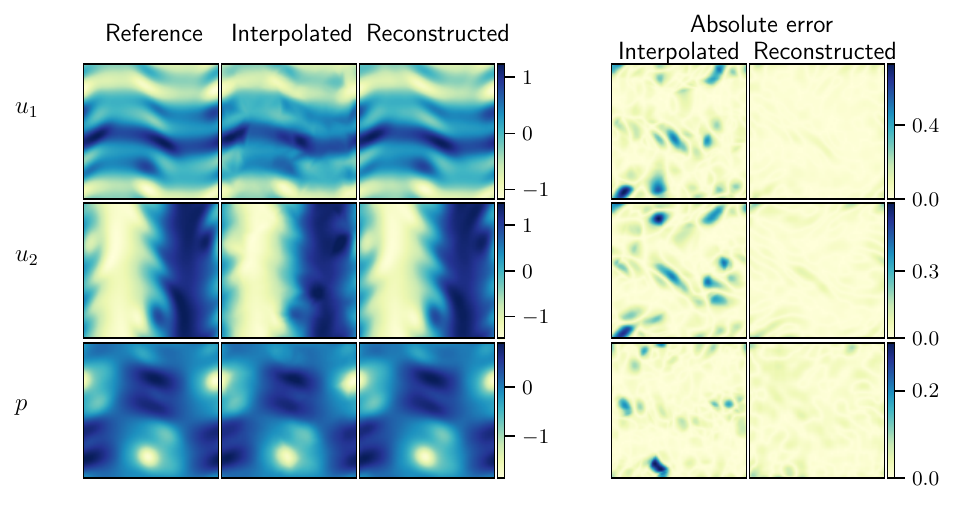}
    \caption{Reconstructed turbulent Kolmogorov flow from non-noisy measurements with the snapshot-enforced loss, at $t=1.7$.}\label{fig:2dkol:clean-snapshots}
\end{figure}

We investigate different loss functions for reconstructing the Kolmogorov flow.
Same as the reconstruction of the laminar wake (Section~\ref{sec:laminar:clean}), all loss functions can achieve reconstruction errors of less than 10\%.
However, unlike the laminar wake, the loss functions have a stronger influence on the reconstruction of the Kolmogorov flow due to its chaotic nature.
Table~\ref{tab:2dkol:clean-compare-lossfn} shows the mean standard deviation of the relative error computed over five trainings with different random weight initialization.
The mean relative error of networks trained with the snapshot-enforced loss $\lossStrict$ is over 20\% lower than the networks trained with $\lossClassic$ and $\lossMean$, while the mean relative errors of the networks trained with $\lossClassic$ and $\lossMean$ are within 3\% of each other.
The standard deviations of the flow reconstructed with loss functions with harder constraints ($\lossStrict$ and $\lossMean$) are approximately half of the flow reconstructed with the softly-constrained loss $\lossClassic$. 
This shows that, as the dynamics become more nonlinear, loss functions with harder constraints on the measurements improve the robustness towards network initialization and the accuracy of the output. 
Figure~\ref{fig:2dkol:clean-snapshots} shows that the reconstructed flow has a smaller instantaneous absolute error.
\begin{table}[htb]
    \centering
    \caption{The snapshot-enforced loss leads to the lowest relative error (mean$\pm$standard~deviation) when reconstructing the turbulent Kolmogorov flow from non-noisy measurements. The relative error is measured over five repeated trainings with different random weight initialization and sensor locations.}\label{tab:2dkol:clean-compare-lossfn}
    \begin{tabular}{|c|c|c|c|c|}
        \hline
         & $\lossClassic$ & $\lossStrict$ & $\lossMean$ & Interpolated \\
        \hline 
        $\epsilon$(\%) & $7.19\pm0.73$ & $5.51\pm0.34$ & $7.38\pm0.33$ & $16.46\pm0.00$ \\
        \hline
    \end{tabular}
\end{table}

\begin{figure}[htb]
    \centering
    \includegraphics[width=5.7in]{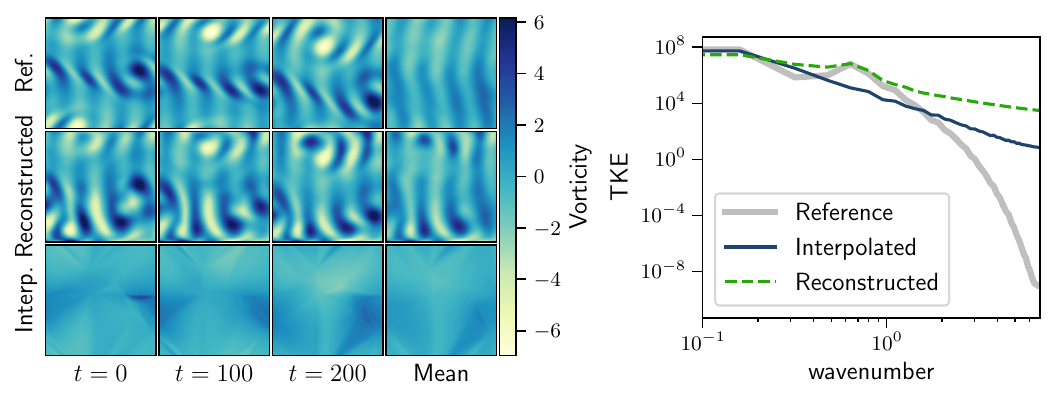}
    \caption{
        The flow reconstructed from ten input sensors and ten general sensors. The reconstruction has recovered the instantaneous vorticity fields and the turbulent kinetic energy spectrum.
    }\label{fig:2dkol:clean-10sensors}
\end{figure}
We also tested how the PC-DualConvNet behaves in extreme situations by randomly removing the sensors in Figure~\ref{fig:2dkol:clean_sensor_locations} until only ten input sensors and ten general sensors remain.
The reconstructed flow using ten sensors has a reconstruction error of 73.1\%, higher than the 69.9\% of the interpolated flow.
However, the instantaneous and mean vorticity field (Figure~\ref{fig:2dkol:clean-10sensors}) shows that the reconstructed flow is physical and shows a closer resemblance to the reference dataset.
The reconstructed flow has also recovered the turbulence energy spectrum better compared to the interpolated flow (Figure~\ref{fig:2dkol:clean-10sensors} right panel).
The results with ten sensors highlight the limitation of measuring the quality of reconstruction using only the $\ell_2$ error.
Even if the interpolated flows have a lower relative error compared to the reconstructed flow, it clearly failed to reconstruct the flow by both the TKE and visual inspection.
There are clear discrepancies between the reconstructed and the reference flow in Figure~\ref{fig:2dkol:clean-10sensors}, showing that more sensors are necessary to achieve a good reconstruction.

\subsection{Reconstructing from noisy measurements}\label{sec:turbulent:noisy}
\begin{figure}[htb]
    \centering
    \begin{subfigure}[c]{0.23\linewidth}
        \includegraphics[width=\linewidth]{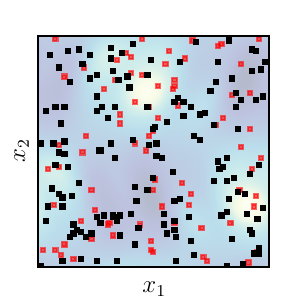}
        \caption{}\label{fig:2dkol:noisy_sensor_locations}
    \end{subfigure}
    \begin{subfigure}[c]{0.7\linewidth}
        \includegraphics[width=\linewidth]{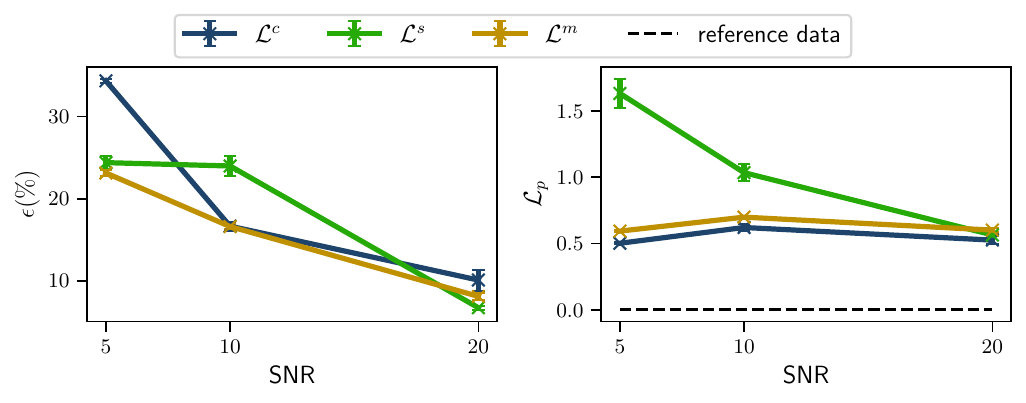}
        \caption{}\label{fig:2dkol:noisy_compare_lossfn}
    \end{subfigure}
    \caption{(a) Sensor placement for reconstructing the turbulent Kolmogorov flows from noisy measurements. (b) Mean and standard deviation of the reconstructed Kolmogorov flow from noisy measurements at different SNR using different loss functions.}
\end{figure}

\begin{figure}
    \centering
    \includegraphics[width=5.5in]{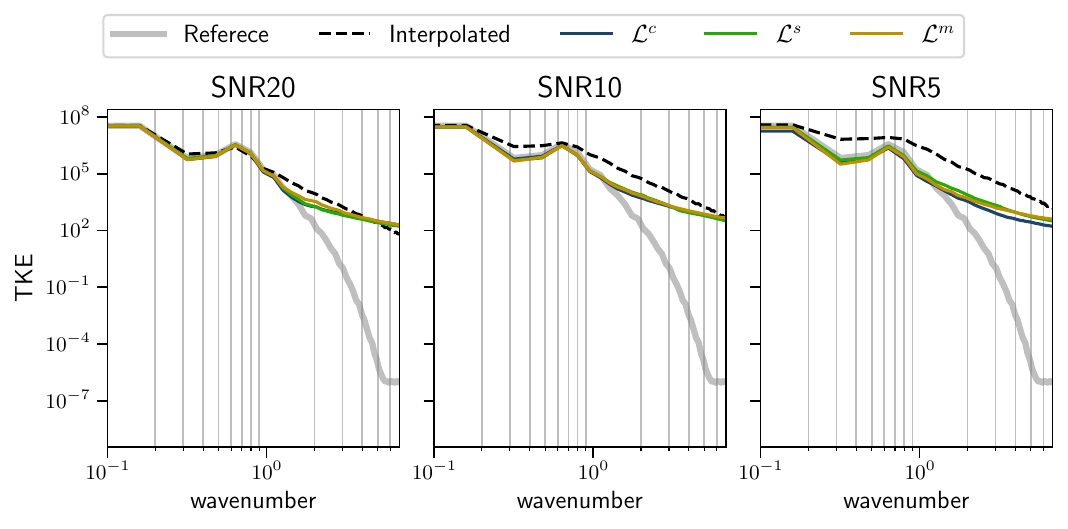}
    \caption{The turbulent kinetic energy (TKE) spectrum. }\label{fig:2dkol:noisy_tke_wavenumbers}
\end{figure}

In this section, we reconstruct the turbulent Kolmogorov flow from sparse measurements corrupted with white noise at $\SNR=20,10\text{ and }5$.
The same number of sensors (80 inputs, 150 general) are placed at random locations, shown in Figure~\ref{fig:2dkol:noisy_sensor_locations}.
The relative errors and the physics loss of the flow reconstructed with $\lossClassic, \lossStrict\text{ and }\lossMean$ are summarized in Figure~\ref{fig:2dkol:noisy_compare_lossfn}.
The physics loss shows a similar trend to Section~\ref{sec:laminar:noisy} --- harder constraints on the measurements lead to higher physics losses as the noise in the measurements increases.
At a low level of noise ($\SNR=20$) $\lossStrict$ leads to the lowest reconstruction error, although all loss functions achieve a relative error lower than 10\%.
Unlike when reconstructing the laminar wake, the softly-constrained loss is able to reconstruct the Kolmogorov flow from measurements at $\SNR=10$, suggesting that some variation of performance is the result of hyperparameter selection when the measurements are corrupted with a medium amount of noise.
However, hyperparameters may be cumbersome to tune, and a (globally) optimal set is rarely found. 
On the other hand, $\lossMean$ has achieved consistent reconstructions of both the laminar flow (Section~\ref{sec:laminar:noisy}) and the turbulent flow from noisy measurements.
As the signal-to-noise ratio decreases further, the reconstruction error of all loss functions increases.
At $\SNR=5$ for $\lossMean$ and $\lossClassic$, the physics loss decreased compared to $\SNR=10$, showing that the reconstruction is favouring satisfying the physics over the noisy measurements.
At $\SNR=5$, the reconstruction error of the flow reconstructed with $\lossClassic$ increased to 34.4\%, higher than both the snapshot-enforced loss and the mean-enforced loss, showing again that harder constraints improve the robustness of the PC-DualConvNet to noise.
The strictly-enforced mean loss achieves a reconstruction error of 23.1\%, similar to the error of the snapshot-enforced loss, but with much lower physics loss.
For all SNRs, the reconstructed flows have recovered the correct energy spectrum (Figure~\ref{fig:2dkol:noisy_tke_wavenumbers}) up to wavenumber $1$, showing that the reconstructed flow is more physical than the interpolated flow even when starting from noisy measurements.
\begin{figure}
    \parbox{\LW{0.1}}{\subcaption{}\label{fig:2dkol:noisy-snapshots20}}\hfill%
    \parbox{\LW{0.9}}{\includegraphics[width=\linewidth]{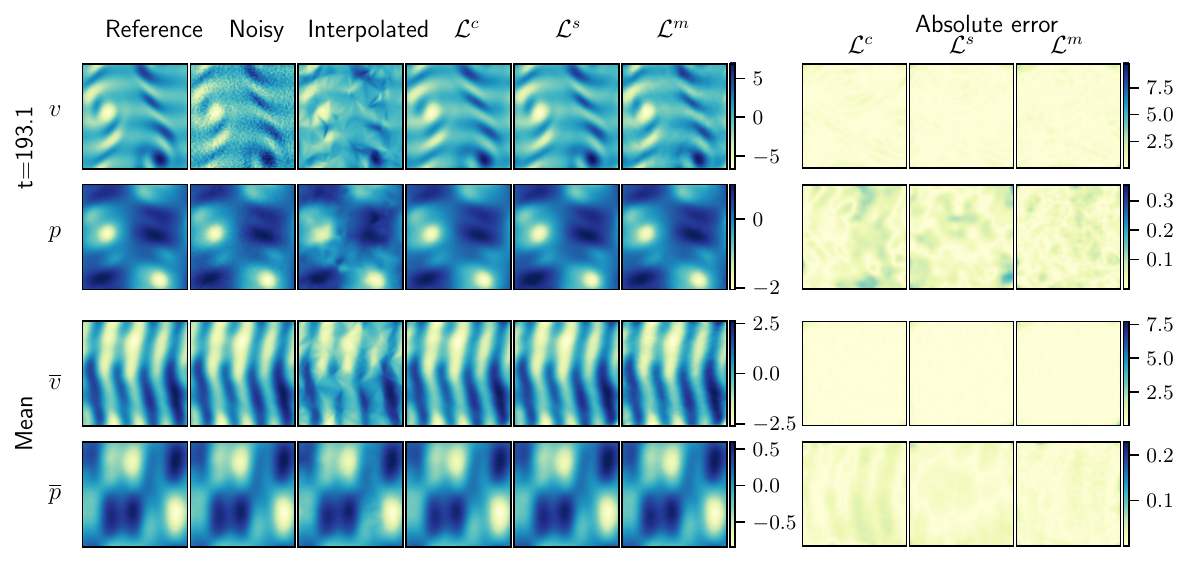}}\hfill%
    \parbox{\LW{0.1}}{\subcaption{}\label{fig:2dkol:noisy-snapshots10}}\hfill%
    \parbox{\LW{0.9}}{\includegraphics[width=\linewidth]{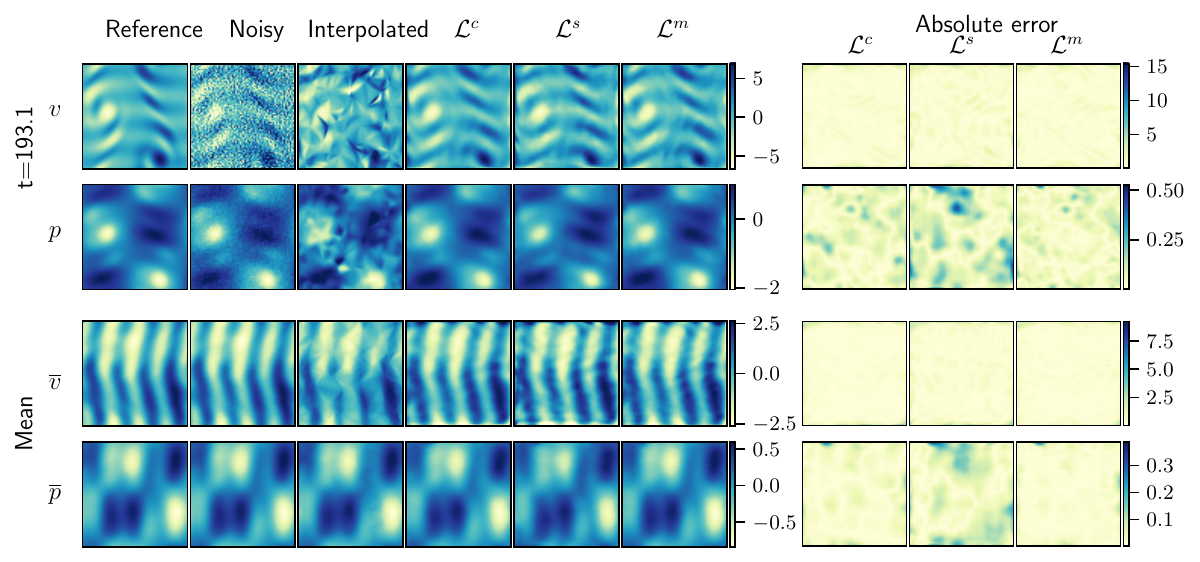}}\hfill%
    \parbox{\LW{0.1}}{\subcaption{}\label{fig:2dkol:noisy-snapshots5}}\hfill%
    \parbox{\LW{0.9}}{\includegraphics[width=\linewidth]{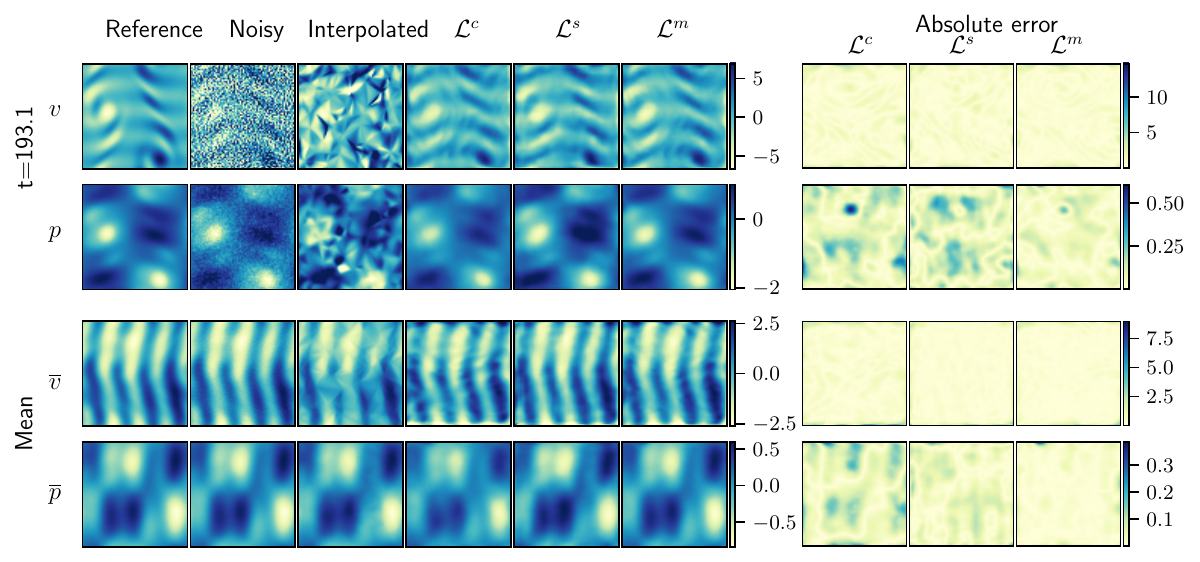}}\hfill%
    \caption{Reconstructed laminar wake from noisy measurements at SNR=20 (\ref{fig:2dkol:noisy-snapshots20}), 10 (\ref{fig:2dkol:noisy-snapshots10}) and 5 (\ref{fig:2dkol:noisy-snapshots5}).}\label{fig:2dkol:noisy-snapshots}
\end{figure}

Figure~\ref{fig:2dkol:noisy-snapshots} compares the instantaneous snapshots and the mean of the reconstructed flows. 
At SNR=20 (Figure~\ref{fig:2dkol:noisy-snapshots20}) there is no visible difference between the reconstructed flows using different loss functions, even though the relative error shows that $\lossClassic$ is slightly worse than others.
The flow reconstructed with $\lossStrict$ starts showing visible noise from SNR=10 (Figure~\ref{fig:2dkol:noisy-snapshots10} and~\ref{fig:2dkol:noisy-snapshots5}). 
At SNR=5, the flow reconstructed with $\lossClassic$ shows a visible reduction in the variation of the flow field, while the flow reconstructed with $\lossMean$ and $\lossStrict$ is not affected.
At all SNRs using any loss function, the instantaneous reconstructed flow shows a closer resemblance to the reference data than the interpolated flow. 
Consider the relative error, the physics loss (Figure~\ref{fig:2dkol:noisy_compare_lossfn}) and the snapshots (Figure~\ref{fig:2dkol:noisy-snapshots}), $\lossMean$ provides the most consistent performance at different levels of noise, providing a small relative error while keeping the physics loss close to the reference physics loss at high levels of noise.
\FloatBarrier%

\section{Conclusion}\label{sec:conclusion}

In this paper, we reconstruct flow fields from sparse, noisy, and heterogeneous measurements (incomplete data) with a physics-constrained dual-branch convolutional neural network (PC-DualConvNet). The network is trained using only incomplete data. 
To recover the missing information and infer the full flow field, the governing equations of the flow are embedded into the loss function.
We investigate three loss functions, i.e., 
the softly-constrained loss, which is common in the flow reconstruction literature~\citep{raissi2019PhysicsinformedNeuralNetworks}; 
the snapshot-enforced loss \citep{gao2021SuperresolutionDenoisingFluid} for steady flows; 
and the mean-enforced loss (proposed here).
First, we reconstruct both the laminar wake and the turbulent Kolmogorov flow from non-noisy measurements from a number of sensors of $\sim 1\%$ of all grid points.
When reconstructing the laminar wake, we find no significant difference between the softly-constrained and the snapshot-enforced losses.
However, when reconstructing the turbulent Kolmogorov, we find that the snapshot-enforced loss leads to a $\sim 25\%$ reduction in the reconstruction error compared to the softly-constrained loss due to the flow being chaotic.
Second, we tested the loss functions on measurements corrupted with white noise at signal-to-noise ratios ($\SNR$) 20, 10 and 5.
We find that the Fourier layer in the PC-DualConvNet improves the robustness of the snapshot-enforced loss to noise, which is designed for non-noisy measurements.
At $\SNR<$10,  the snapshot-enforced loss leads to noisy reconstructed flows because the hard constraint on the instantaneous measurements prevents the network from denoising.
On the other hand, the softly-constrained loss leads to a smaller variability in the reconstructed flow dynamics because the network favours minimising the physics loss over the data loss.
Finally, combining the benefit of the softly-constrained and the snapshot-enforced losses, we propose the mean-enforced loss.
At large levels of noise, the flow reconstructed with the mean-enforced loss is physical and does not spuriously attenuate the flow variations, while achieving similar reconstruction errors to the snapshot-enforced loss.
The reconstruction errors show that hard constraints in the loss functions make the PC-DualConvNet more generalisable.
In conclusion, we find the snapshot-enforced loss is suitable for reconstructing flows from non-noisy measurements, and the mean-enforced loss is suitable for reconstructing flows in more general situations, i.e., when the measurements are noisy.

\section*{Acknowledgement}
We acknowledge funding from the Engineering and
Physical Sciences Research Council, UK and financial support from the
ERC Starting Grant PhyCo 949388. L.M. is also grateful for the support
from the grant EU-PNRR YoungResearcher TWIN ERC-PI\_0000005. 
Y.M. acknowledges support from the Department of Aeronautics, Imperial College.

\section*{Code Availability}
Codes are available at \url{https://github.com/MagriLab/FlowReconstructionFromExperiment}.

\clearpage
\bibliography{References}

\clearpage%
\appendix
\section{Appendix} 
\subsection{Grid sensitivity study for the reconstruction of the Kolmogorov flow}\label{sec:appendix:2dkol-grid-sensitivity}
\begin{figure}[htb]
    \centering
    \includegraphics[width=\linewidth]{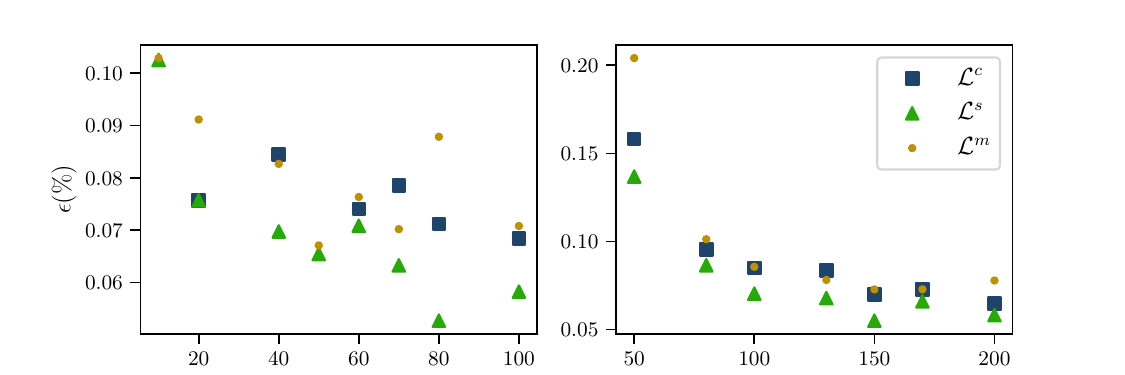}
    \caption{Reconstruction error in response to changing the number of (left) pressure inputs $\mat{P}_{in}$ while holding the number of sensors constant at 200; and (right) sensors $\mat{D}(\vector{x}_s)$ while holding the number of pressure inputs constant at 80. This grid sensitivity test is performed on a Kolmogorov flow simulated with a different initial random seed.}\label{fig:2dkol:clean-grid-sensitivity}
\end{figure}

We use a Kolmogorov flow dataset generated with a different initial random seed than the dataset used in the rest of Section~\ref{sec:turbulent} to determine the number of sensors to place in the domain.
A different random seed is used to eliminate possible bias in sensor placement.
In order words, we want to make sure the reconstruction results presented in the main text come from a previously unseen dataset.
We followed the procedure below to determine the number of sensors:
\begin{enumerate}
    \item Conduct a hyperparameter search for all loss functions $\lossClassic, \lossStrict \text{ and }\lossMean$ using 80 input sensors ($\vector{x}_{in}$) and 200 general sensors ($\vector{x}_s$) randomly placed in the domain.
    \item Randomly remove the input pressure sensors while holding the number of general sensors constant at 200, until the relative error starts to increase (Figure~\ref{fig:2dkol:clean-grid-sensitivity} left panel). We find that the relative error of the reconstructed flow starts to increase when the number of input sensors is smaller than 80.
    \item Randomly remove the general sensors while holding the number of input sensors constant at 80, until the relative error starts to increase (Figure~\ref{fig:2dkol:clean-grid-sensitivity} right panel). We find that the relative error starts to increase when using fewer than 150 sensors.
\end{enumerate}
For Section~\ref{sec:turbulent:clean} and~\ref{sec:turbulent:noisy}, we use 80 input sensors and 150 general sensors randomly placed in the domain to reconstruct the flow.

\FloatBarrier
\subsection{Model structures and hyperparameters}
\begin{figure}[H]
    \centering
    \includegraphics[width=6.5in]{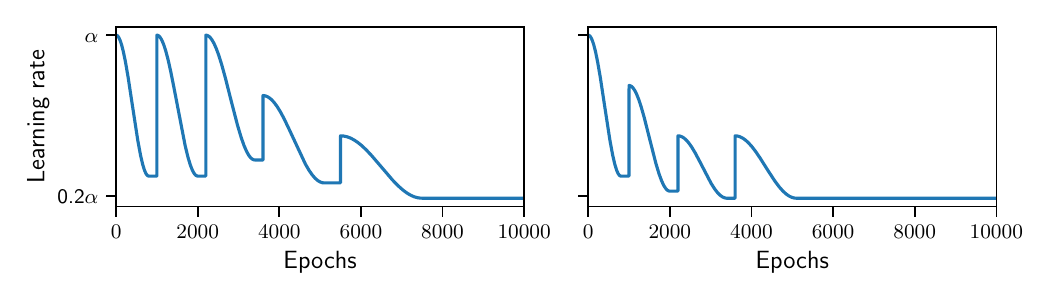}
    \caption{The cyclic decay learning rate schedule applied to reconstruct the laminar flow (left) and the Kolmogorov flow (right), where $\alpha$ is the initial learning rate.}\label{fig:ap:laminar:learning-rate-schedule}
\end{figure}

This section contains the training- and network-related parameters for all examples shown in this paper. 
Figure~\ref{fig:ap:laminar:learning-rate-schedule} shows the learning rate schedules employed in the reconstruction of the laminar wake and the Kolmogorov flow. 
Table~\ref{tab:ap:laminar:clean-params} to~\ref{tab:ap:laminar:noisy-params-mean} contains the network and training parameters for the laminar wake test cases.
Table~\ref{tab:ap:2dkol:params-classic} to~\ref{tab:ap:2dkol:params-mean} contains the network and training parameters for the Kolmogorov flow test cases. 
The parameter `bottleneck image dimension' refers to the smallest image dimension achieved through the resizing layers in the lower branch of the PC-DualConvNet.
\begin{table}[H]
    \centering
    \begin{tabular}{|c|c|}
        \hline
        Bottleneck image dimension & (64,16) \\
        Convolution filter size (lower branch) & (5,5) \\
        Convolution filter size (all other branches) & (3,3) \\
        Input branch channels & [3,] \\
        Upper branch channels & [8,] \\
        Lower branch channels & [8,16,8] \\
        Output branch channels & [4,3] \\
        FFT & off \\
        Batch size & [35] \\
        $\mathbf{\lambda_{div}}$ & 1.0 \\
        $\mathbf{\lambda_{mom}}$ & 1.0 \\
        $\mathbf{\lambda_o}$ & 40.0 \\
        Initial learning rate ($\alpha$) & 0.0023 \\
        Regularisation & 0.0 \\
        Dropout rate & 0.0 \\
        \hline
    \end{tabular}
    \caption{The network structure and hyperparameters used to reconstruct the laminar wake from non-noisy measurements.}\label{tab:ap:laminar:clean-params}
\end{table} 
\begin{table}[H]
    \centering
    \begin{tabular}{|c|c|c|c|}
        \hline
         & SNR=20 & SNR=10 & SNR=5\\
        \hline
        Upper branch channels & [4,4] & [4,4] & [4,4]\\
        Initial learning rate ($\alpha$) & 0.0013 & 0.0023 & 0.0012 \\
        Batch size & 46 & 67 & 67 \\
        Regularisation & 0.007 & 0.05 & 0.022\\
        $\mathbf{\lambda_{div}}$ & 2.0 & 2.1 & 3.8\\
        $\mathbf{\lambda_o}$ & 12.0 & 11.0 & 11.0\\
        Dropout rate & 0.05\% & 0.5\% & 0.5\% \\
        \hline
    \end{tabular}
    \caption{The structure and hyperparameters of the network using the softly-constrained loss, from noisy measurements. Showing the difference from~\ref{tab:ap:laminar:clean-params} only.}\label{tab:ap:laminar:noisy-params-classic}
\end{table} 

\begin{table}[H]
    \centering
    \begin{tabular}{|c|c|c|c|}
        \hline
         & SNR=20 & SNR=10 & SNR=5\\
        \hline
        Upper branch channels & [1,] & [1,] & [1,]\\
        FFT & on & on & on \\
        Initial learning rate ($\alpha$) & 0.014 & 0.0068 & 0.0088 \\
        Batch size & 37 & 46 & 46 \\
        Regularisation & 0.095 & 0.056 & 0.070\\
        $\mathbf{\lambda_{div}}$ & 1.0 & 1.0 & 1.0\\
        $\mathbf{\lambda_o}$ & n/a & n/a & n/a\\
        Dropout rate & 0.48\% & 0.55\% & 0.6\% \\
        \hline
    \end{tabular}
    \caption{The structure and hyperparameters of the network using the snapshot-enforced loss, from noisy measurements. Showing the difference from~\ref{tab:ap:laminar:clean-params} only.}\label{tab:ap:laminar:noisy-params-strict}
\end{table}
\begin{table}[H]
    \centering
    \begin{tabular}{|c|c|c|c|}
        \hline
         & SNR=20 & SNR=10 & SNR=5\\
        \hline
        Upper branch channels & [1,] & [4,4] & [8,]\\
        Initial learning rate ($\alpha$) & 0.003 & 0.0023 & 0.0046 \\
        Batch size & 50 & 200 & 120 \\
        Regularisation & 0.076 & 0.09 & 0.042 \\
        $\mathbf{\lambda_{div}}$ & 2.5 & 2.7 & 2.6 \\
        $\mathbf{\lambda_o}$ & 31.5 & 48.5 & 24.0\\
        Dropout rate & 0.22\% & 0.25\% & 0.97\% \\
        \hline
    \end{tabular}
    \caption{The structure and hyperparameters of the network using the mean-enforced loss, from noisy measurements. Showing the difference from~\ref{tab:ap:laminar:clean-params} only.}\label{tab:ap:laminar:noisy-params-mean}
\end{table}
\begin{table}[H]
    \centering
    \begin{tabular}{|c|c|c|c|c|}
        \hline 
         & Non-noisy & SNR=20 & SNR=10 & SNR=5 \\
        \hline
        Bottleneck image dimension & (16,16) & & & \\
        Convolution filter size (upper branches) & (5,5) & & & \\
        Convolution filter size (lower branches) & (5,5) & & (3,3) & (3,3) \\
        Convolution filter size (all other) & (3,3) & & & \\
        Input branch channels & [3,] & & & \\
        Upper branch channels & [4,4] & & & \\
        Lower branch channels & [4,8,16,8,4] & & & \\
        Output branch channels & [4,3] & & & \\
        FFT & off & & & \\
        Batch size & 182 & & 214 & 240\\
        $\mathbf{\lambda_{div}}$ & 1.0 & & & \\
        $\mathbf{\lambda_{mom}}$ & 1.0 & & & \\
        $\mathbf{\lambda_o}$ & 8.0 & & 26.1 & 7.6\\
        Initial learning rate ($\alpha$) & 0.004 & & 0.00034 & 0.00018\\
        Regularisation & 0.0 & & 0.0012 & 0.0037 \\
        Dropout rate & 0.0\% & & 0.24\% & 0.61\% \\
        \hline
    \end{tabular}
    \caption{The structure and hyperparameters of the network used to reconstruct the Kolmogorov flow with softly-constrained loss. The empty cells in column 2~through~4 represent parameters that are the same as in column~1.}\label{tab:ap:2dkol:params-classic}
\end{table}
\begin{table}[H]
    \centering
    \begin{tabular}{|c|c|c|c|c|}
        \hline 
         & Non-noisy & SNR=20 & SNR=10 & SNR=5 \\
        \hline
        Bottleneck image dimension & (16,16) & & & \\
        Convolution filter size (upper branch) & (5,5) & & & \\
        Convolution filter size (lower branch) & (5,5) & & & (3,3) \\
        Convolution filter size (all other) & (3,3) & & & \\
        Input branch channels & [3,] & & & \\
        Upper branch channels & [1,] & & & \\
        Lower branch channels & [4,8,16,8,4] & & & \\
        Output branch channels & [4,3] & & & \\
        FFT & on & & & \\
        Batch size & 158 &  & 214 & 214\\
        $\mathbf{\lambda_{div}}$ & 1.0 & & & \\
        $\mathbf{\lambda_{mom}}$ & 1.0 & & & \\
        Initial learning rate ($\alpha$) & 0.0008 & & 0.0028 & 0.0088 \\
        Regularisation & 0.0 & & 0.008 & 0.0056 \\
        Dropout rate & 0.0\% & & 0.59\% & 0.17\% \\
        \hline
    \end{tabular}
    \caption{The structure and hyperparameters of the network used to reconstruct the Kolmogorov flow with snapshot-enforced loss. The empty cells in column 2~through~4 represent parameters that are the same as in column~1.}\label{tab:ap:2dkol:params-strict}
\end{table}
\begin{table}[H]
    \centering
    \begin{tabular}{|c|c|c|c|c|}
        \hline 
         & Non-noisy & SNR=20 & SNR=10 & SNR=5 \\
        \hline
        Bottleneck image dimension & (16,16) & & & \\
        Convolution filter size (upper branch) & (5,5) & & & \\
        Convolution filter size (lower branch) & (5,5) & & (3,3) & (3,3)\\
        Convolution filter size (all other) & (3,3) & & & \\
        Input branch channels & [3,] & & & \\
        Upper branch channels & [4,] & & & \\
        Lower branch channels & [4,8,16,8,4] & & & \\
        Output branch channels & [4,3] & & & \\
        FFT & off & & & \\
        Batch size & 600 & & 857 & 1200 \\
        $\mathbf{\lambda_{div}}$ & 1.0 & & & \\
        $\mathbf{\lambda_{mom}}$ & 1.0 & & & \\
        $\mathbf{\lambda_o}$ & 64.0 & & 34.6 & 5.9 \\
        Initial learning rate ($\alpha$) & 0.0047 & & 0.0013 & 0.0013 \\
        Regularisation & 0.0 & & 0.0007 & 0.0039\\
        Dropout rate & 0.0\% & & 0.21\% & 0.23\% \\
        \hline
    \end{tabular}
    \caption{The structure and hyperparameters of the network used to reconstruct the Kolmogorov flow with mean-enforced loss. The empty cells in column 2~through~4 represent parameters that are the same as in column~1.}\label{tab:ap:2dkol:params-mean}
\end{table}

\subsection{Learning curves}
\begin{figure}[H]
    \centering
    \includegraphics[width=\linewidth]{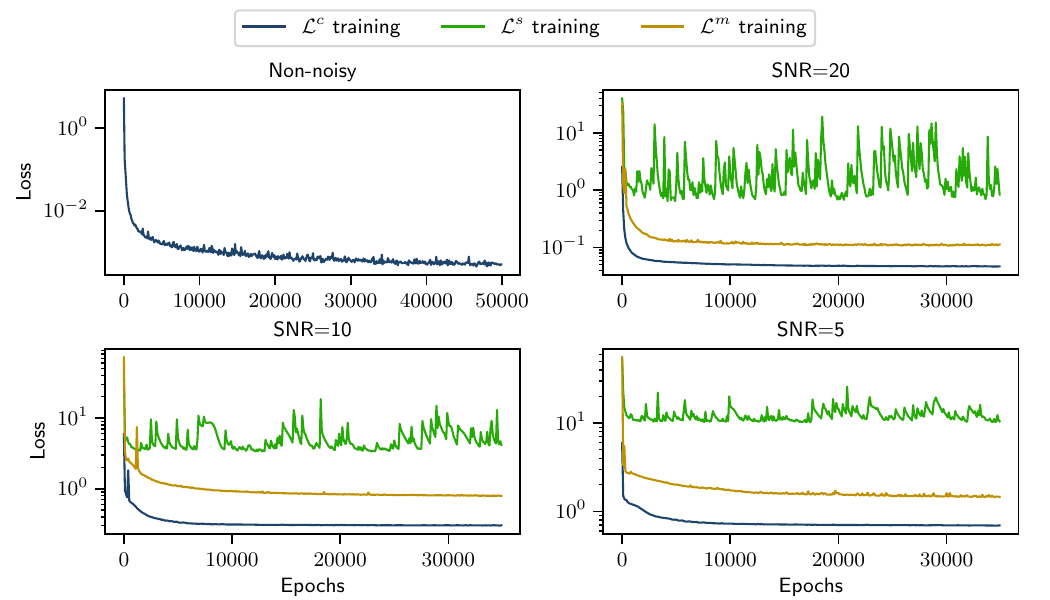}
    \caption{Learning curves of the reconstruction of the laminar wake dataset. Learning curves are shown for the test runs plotted in Figure~\ref{fig:2dtriangle:clean} and~\ref{fig:2dtriangle:noisy-snapshots}.}
    \label{fig:ap:loss-2dtriangle}
\end{figure}

\begin{figure}[H]
    \centering
    \includegraphics[width=\linewidth]{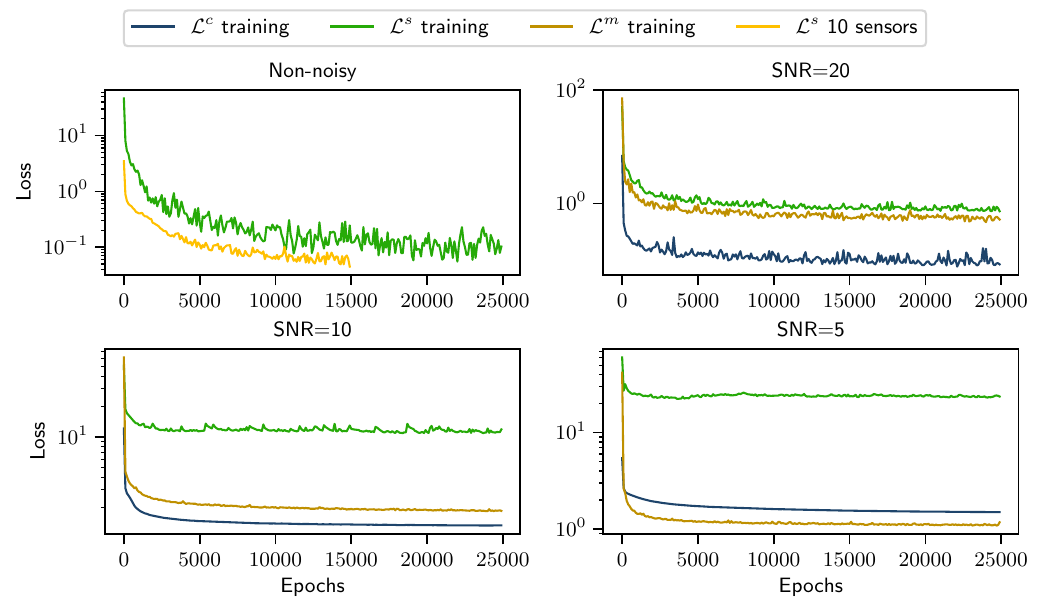}
    \caption{Learning curves of the reconstruction of the turbulent Kolmogorov flow. Learning curves are shown for the test runs plotted in Figure~\ref{fig:2dkol:clean-snapshots}, \ref{fig:2dkol:clean-10sensors} and~\ref{fig:2dkol:noisy-snapshots}.}
    \label{fig:ap:loss-2dkol}
\end{figure}

Figure~\ref{fig:ap:loss-2dtriangle} and~\ref{fig:ap:loss-2dkol} show the learning curves for the laminar wake and Kolmogorov examples shown in this paper.

\end{document}